\documentclass[aps,reprint,groupaddress,superscriptaddress,amssymb,prx]{revtex4-2}
\usepackage[utf8]{inputenc}
\usepackage{color}
\usepackage{xcolor}
\usepackage{graphicx}
\usepackage{stmaryrd}

\usepackage{mdframed}
\usepackage{physics}
\usepackage{amsmath}
\usepackage{amsfonts}
\usepackage{bm}
\usepackage[symbol]{footmisc}

\begin{document}
\preprint{APS/123-QED}

\title{Information interference driven by environmental activity}

\author{Giorgio Nicoletti}
\affiliation{ECHO Laboratory, École Polytechnique Fédérale de Lausanne, Lausanne, Switzerland}
\author{Daniel Maria Busiello}
\affiliation{Max Planck Institute for the Physics of Complex Systems, Dresden, Germany}

\begin{abstract}
\noindent Real-world systems are shaped by both their complex internal interactions and the changes in their noisy environments. In this work, we study how a shared active bath affects the statistical dependencies between two interacting Brownian particles by evaluating their mutual information. We decompose the mutual information into three terms: information stemming from the internal interactions between the particles; information induced by the shared bath, which encodes environmental changes; a term describing information interference that quantifies how the combined presence of both internal interactions and environment either masks (destructive interference) or boosts (constructive interference) information. By studying exactly the case of linear interactions, we find that the sign of information interference depends solely on that of the internal coupling. However, when internal interactions are described by a nonlinear activation function, we show that both constructive and destructive interference appear depending on the interplay between the timescale of the active environment, the internal interactions, and the environmental coupling. Finally, we show that our results generalize to hierarchical systems where asymmetric couplings to the environment mimic the scenario where the active bath is only partially accessible to one particle. This setting allows us to quantify how this asymmetry drives information interference. Our work underscores how information and functional relationships in complex multi-scale systems are fundamentally shaped by the environmental context.
\end{abstract}

\maketitle
    
\vspace{0.5cm}

\section{Introduction}
\noindent The interplay between internal and extrinsic processes is a key driver in many real-world systems, which are often coupled to noisy, ever-changing environments that cannot be directly observed. The importance of environmental variability has been underscored in several studies, ranging from biological and ecological systems \cite{Hilfinger2011, Tsimring2014, Zhu2009, wienand2017fluctuating, wienand2018eco, padmanabha2024spatially}, to biochemical and regulatory networks \cite{swain2002extrinsic, thomas2014phenotypic, bowsher2012biochemical, dass2021equilibrium}, to diffusion in crowded or inhomogeneous media \cite{dorsaz2010diffusion, galanti2014diffusion, chechkin2017brownian}. Moreover, emergent features in biological contexts, such as thermophoresis \cite{piazza2008thermophoresis, liang2022emergent}, chemical selection \cite{busiello2021dissipation, berton2020thermodynamics, liang2024thermodynamic}, and pattern formation \cite{falasco2018turing, brauns2020phase}, have been recently shown to be sheer consequences of the interplay between environmental and internal interactions acting on different timescales, typically in nonequilibrium conditions \cite{busiello2020coarsegrained}.

Disentangling the effect of the environment and its nonequilibrium activity from that of internal couplings is usually a challenging task. Recent information-theoretic approaches \cite{nicoletti2021mutual,nicoletti2022mutual} started tackling this problem by analyzing the mutual information between observable degrees of freedom induced by a shared fluctuating environment. Although in some cases this disentangling is exact \cite{nicoletti2021mutual}, non-trivial behaviors of the mutual information can appear, e.g., in the presence of nonlinear interactions, and are analytically captured by an interference term \cite{nicoletti2022mutual}. Nevertheless, the role of nonequilibrium activity remains unclear in this picture.

Countless systems from biology, e.g., bacteria \cite{elgeti2015physics,wu2000particle}, enzymes \cite{ghosh2021enzymes}, and red blood cells \cite{di2024variance}, are important examples of active matter that is maintained out of equilibrium by internal or external energy consuming processes \cite{fodor2022irreversibility,julicher2018hydrodynamic}. In numerous theoretical and experimental scenarios, such as passive tracers probing the surrounding media \cite{banerjee2022tracer} and fluorescent molecules in a chemical buffer \cite{xu2019direct}, active systems act as environments for other observed components \cite{maes2020fluctuating,dabelow2019irreversibility}. In general, how this environmental activity affects functional couplings and information propagation is currently unknown.


In this work, we study how information between two interacting particles is shaped by the presence of a shared, active bath, which acts as a common environment. We model the action of the bath as a colored noise with a given auto-correlation time. Since the interplay between fast and slow processes is known to be a pivotal feature of biological, chemical, and ecological systems \cite{hastings2010timescales, poisot2015beyond, timme2014multiplex, cavanagh2020diversity, zeraati2023intrinsic, nicoletti2023emergence, nicoletti2024information}, we explore the behavior of information at different environmental timescales, as well as different values of both environmental and internal couplings. To do so, we decompose the mutual information between the two particles into three distinct contributions \cite{nicoletti2022mutual}: the internal information that is solely due to internal couplings; the environmental information induced by the coupling of both particles with the active bath; and an information interference term modeling the combined effect of internal interactions and active environment taken together. Although information interference has been introduced in nonlinear scenarios, the presence of an active bath generally triggers its emergence even for simple linearized dynamics. In particular, we compute the interference term both in the presence of linear internal interactions and with nonlinear activation functions, showing that both constructive and destructive information interference may occur. The active environment may drive or suppress the information between the two particles depending on both the bath-particle couplings and the timescales at play. Furthermore, we uncover a similar phenomenology also when the bath is coupled to each particle in a different way. This setup mimics the fact that the active bath may act as a partially hidden degree of freedom, being only marginally accessible to one of the particles. Our results clearly show that active environments play a key role in shaping the statistical dependencies in a system, either masking or boosting the information between the particles they influence.

\section{Information between interacting particles in an active environment}
\noindent We consider a general model of two interacting particles, $x$ and $y$, evolving in an active bath:
\begin{equation}
\label{eqn:model}
    \begin{gathered}
        \tau \dot{x}(t) = -x(t) + g F_x(x, y) + \sqrt{2 D_x \tau} \xi_x(t) + \gamma \eta(t) \\
        \tau \dot{y}(t) = -y(t) + g F_y(x, y) + \sqrt{2 D_y \tau} \xi_y(t) + \gamma \eta(t) \;.
    \end{gathered}
\end{equation}
Here, $\tau$ sets the timescale of the evolution of the two particles, $F_x$ and $F_y$ are generic force fields, $g$ determines the interaction strength, $D_x$ and $D_y$ are the diffusion coefficients, and $\xi_x$, $\xi_y$ uncorrelated Gaussian noises,
\begin{equation}
    \ev{\xi_x(t)} = \ev{\xi_y(t)} = 0, \quad \ev{\xi_x(t)\xi_y(t')} = \delta_{xy}\delta(t - t') \; .
\end{equation}
Both particles interact with an active bath $\eta$, through a coupling parameter $\gamma$. We model the action of the bath as a colored noise
\begin{equation}
    \ev{\eta(t)} = 0, \quad \ev{\eta(t)\eta(t')} = D_\eta \exp\left[-\frac{|t - t'|}{\tau_\eta}\right]
\end{equation}
where $D_\eta$ controls the amplitude of the active noise, and $\tau_\eta$ is its timescale. As such, $\eta$ can be interpreted as a fictitious variable following a linear Langevin equation with drift $-1/\tau_\eta$ and noise coefficient $\sqrt{2 D_\eta/\tau_\eta}$. This corresponds to the customary active Ornstein-Uhlenbeck model to describe active baths \cite{dabelow2019irreversibility}.

We are interested in understanding how the coupling to the active environment $\gamma$ influences the effective interaction between the internal degrees of freedom, i.e., the two particles immersed in the bath. To do so, we evaluate the mutual information between $x$ and $y$,
\begin{equation}
\label{eqn:Ixy}
    I_{xy} = \int dxdy\,p_{xy}^\mathrm{st}(x,y) \log \frac{p_{xy}^\mathrm{st}(x,y)}{p_{x}^\mathrm{st}(x)p_{y}^\mathrm{st}(y)}
\end{equation}
where $p_{xy}^\mathrm{st}(x,y)$ is the stationary joint probability distribution of the particles, and $p_{x}^\mathrm{st}$, $p_{y}^\mathrm{st}$ the corresponding marginal distributions. Eq.~\eqref{eqn:Ixy} quantifies the statistical dependency between $x$ and $y$ in terms of how much information is contained in their joint distribution with respect to its factorization. In particular, $I_{xy} = 0$ if and only if the two particles are independent. On the other hand, a large value of the mutual information signals that their effective coupling - in terms of their statistical interdependence - is strong.

Clearly, in the absence of the active environment, the mutual information $I_{xy}$ receives contributions from the force fields $F_x$ and $F_y$. Hence, it is useful to introduce the internal mutual information,
\begin{equation}
\label{eqn:Ixy_int}
    I_{xy}^\mathrm{int} = \int dxdy\,p_{xy}^\mathrm{st}(\gamma = 0) \log \frac{p_{xy}^\mathrm{st}(\gamma = 0)}{p_{x}^\mathrm{st}(\gamma = 0)p_{y}^\mathrm{st}(\gamma = 0)}
\end{equation}
where we omitted, for brevity, the dependency of the probability distributions on $x$ and $y$. Here, $(\gamma = 0)$ indicates that this mutual information has to be computed as if the environment were absent, i.e., by using Eq.~\eqref{eqn:model} with $\gamma$ set to $0$. Eq.~\eqref{eqn:Ixy_int} quantifies the information between $x$ and $y$ generated solely by their internal interactions, i.e., $F_x$ and $F_y$ in our case. However, as we will see, this kind of interaction is not the sole source of statistical dependencies in the complete system described by Eq.~\eqref{eqn:model}.

\subsection{Information from an active environment}
\noindent To first understand the baseline effect of the environment, we now consider the case of two non-interacting particles, i.e., we set $g = 0$. Incorporating the dynamics of $\eta$ into Eq.~\eqref{eqn:model}, in this simplified setting the model describes a three-dimensional Ornstein-Uhlenbeck process whose stationary covariance matrix $\hat{\Sigma}$ obeys the Lyapunov equation
\begin{equation}
\label{eqn:lyap}
    \hat{A} \hat{\Sigma} + \hat{\Sigma}\hat{A}^T = 2 \hat{D}
\end{equation}
where $\hat{A}$ is the interaction matrix,
\begin{equation*}
    \hat{A} = \begin{pmatrix}
        1/\tau & 0 & -\gamma/\tau \\
        0 & 1/\tau & -\gamma/\tau \\
        0 & 0 & 1/\tau_\eta \\
    \end{pmatrix} \; ,
\end{equation*}
and $\hat{D} = \text{diag}(D_x/\tau, D_y/\tau, D_\eta/\tau_\eta)$ is the diffusion matrix of the particles and $\eta$ together, i.e., of the vector $(x, y, \eta)$. The stationary solution for $x$ and $y$ is the following Gaussian distribution
\begin{equation*}
    p_{xy}^\mathrm{env} = \mathcal{N}(0, \hat{\Sigma}^\mathrm{env}) \;,
\end{equation*}
where the superscript denotes the fact that only the coupling between the environment and the internal degrees of freedom is present. In particular, $\hat{\Sigma}^\mathrm{env}$ is the submatrix of $\hat{\Sigma}$ whose rows and columns correspond to $x$ and $y$ and is given by
\begin{equation}
    \hat{\Sigma}^\mathrm{env} = \begin{pmatrix}
        D_x + \frac{D_\gamma}{1 + \alpha^{-1}} & \frac{D_\gamma}{1 + \alpha^{-1}} \\
        \frac{D_\gamma}{1 + \alpha^{-1}} & D_y + \frac{D_\gamma}{1 + \alpha^{-1}}\\ 
    \end{pmatrix}
\end{equation}
where we introduced the dimensionless parameters $D_\gamma = D_\eta \gamma^2$ and $\alpha = \tau_\eta / \tau$. It is worth noticing that $\alpha$ controls the timescale separation between the active bath and the particle dynamics, while $D_\gamma$ is a global parameter encoding the strength of the bath-particle interaction.

In this scenario, the mutual information between $x$ and $y$ then is given by
\begin{align}
\label{eqn:Ixy_env}
    I_{xy}^\mathrm{env} & = \log\sqrt{\frac{\hat{\Sigma}_{xx}^\mathrm{env} \hat{\Sigma}_{yy}^\mathrm{env}}{\det \hat{\Sigma}^\mathrm{env}}} = \\
    & = \frac{1}{2} \log\frac{\left[(D_x + (D_x + D_\gamma) \alpha\right]\left[ D_y + (D_y + 
        D_\gamma) \alpha)\right]}{(1 + \alpha) \left[D_y D_\gamma \alpha +
      D_x (D_y + (D_y + D_\gamma) \alpha)\right]} \nonumber \; .
\end{align}
A non-zero mutual information between the non-interacting particles demonstrates that the shared active environment is generating an effective coupling between them, similar to what has been observed in previous studies on different systems \cite{nicoletti2021mutual, nicoletti2022mutual}. In particular, we have 
\begin{equation}
\label{eqn:Ienv_limits}
\begin{gathered}
    \lim_{\alpha \to 0} I_{xy}^\mathrm{env} = 0 \;, \\
    \lim_{\alpha \to \infty} I_{xy}^\mathrm{env} = \frac{1}{2} \log\frac{(D_x + D_\gamma)(D_y + D_\gamma)}{D_\gamma(D_x + D_y) + D_x D_y} > 0 \;.
\end{gathered}
\end{equation}
As a consequence, the information induced on the internal degrees of freedom, $x$ and $y$, vanishes if the environment evolves on a faster timescale, i.e., $\alpha \to 0$, consistently with previous results \cite{nicoletti2024information, nicoletti2024gaussian}. Nevertheless, since $\partial_\alpha I_{xy}^\mathrm{env} \ge 0$, the slower the dynamics of the shared environment, the larger the resulting dependency between $x$ and $y$ is going to be, up to the limiting value reached for $\alpha \to \infty$, i.e., for an infinitely slower environment. As expected, Eq.~\eqref{eqn:Ixy_env} diverges when $D_\gamma \to \infty$, indicating that the coupling of the particles to the bath can increase such information indefinitely, in principle.

\subsection{Information interference}
\noindent Hence, as shown in the previous subsection, two non-interacting particles may share information due to the common influence of the active environment. However, when $g$ is different from zero, the internal interactions will also contribute to $I_{xy}$. However, the combined effect of internal interactions and environmental features might be highly complex, potentially hindering or enhancing the mutual information between the particles. Following \cite{nicoletti2022mutual}, to capture how this interplay shapes information, we decompose $I_{xy}$ as follows:
\begin{equation}
    \label{eqn:interference}
    I_{xy} = I_{xy}^\mathrm{int} + I_{xy}^\mathrm{env} + \Xi_{xy} \;.
\end{equation}
The first two terms in Eq.~\eqref{eqn:interference} are always positive and respectively represent the limiting scenarios where the environment is absent, Eq.~\eqref{eqn:Ixy_int}, and where there are no internal interaction, Eq.~\eqref{eqn:Ixy_env}. As such, their sum represents the information between $x$ and $y$ stemming from their internal interactions and their shared coupling on the active environment, independently. The term $\Xi_{xy}$, instead, quantifies the deviations from this baseline due to the simultaneous presence of both the environment and the interactions. In general, this is not a mutual information, i.e., it can be either positive or negative. If $\Xi_{xy} < 0$, the total information $I_{xy}$ is smaller than the sum of that coming from internal interactions and the environment separately, meaning that the combined effect of these two sources hinders information. Conversely, when $\Xi_{xy} > 0$, the dependency between the particle is amplified by the presence of both interactions and the environment. For these reasons, $\Xi_{xy}$ is referred to as the term modeling information interference between the two contributions, which can be either constructive - $\Xi_{xy} > 0$ - or destructive - $\Xi_{xy} < 0$.

Since $I_{xy}$ must be positive, we immediately have that the amount of destructive interference is bounded by $|\Xi_{xy}| \le I_{xy}^\mathrm{int} + I_{xy}^\mathrm{env}$ for $\Xi_{xy} < 0$, with the equality being achieved in the case of total destructive interference. In this case, the information from the internal interactions and that from the environment perfectly counterbalance and cancel each other out, leaving the two particles statistically independent.


\section{Activity-driven information interference in harmonic potentials}
\begin{figure*}
    \centering
    \includegraphics[width=\textwidth]{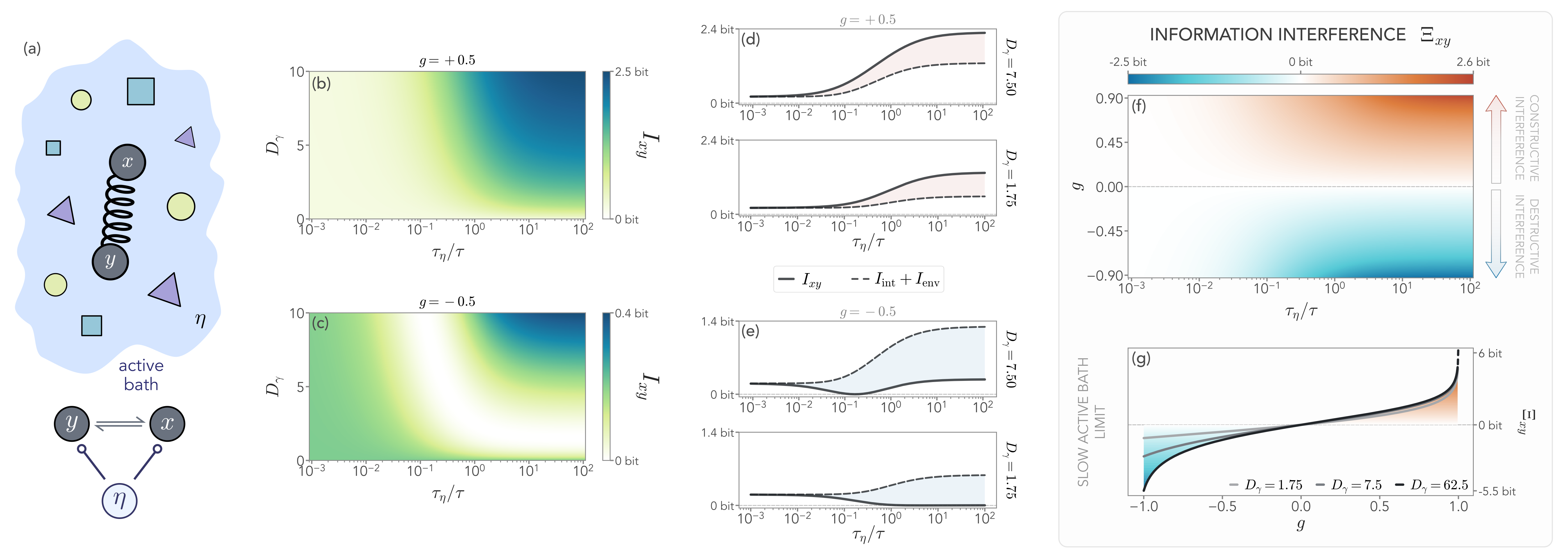}
    \caption{(a) Sketch of the model, describing two interactive particles $x$ and $y$ in an active bath. (b) In a harmonic potential with a positive coupling $g = 0.5$, the mutual information between $x$ and $y$ increases both with the timescale of the bath, $\tau_\eta /\tau$, and with the dimensionless bath coupling, $D_\gamma = \gamma^2 D_\eta$. (c) When the coupling is negative, $g = -0.5$, $I_{xy}$ becomes non-monotonic in $\tau_\eta /\tau$, showing a band of total destructive interference. (d-e) For positive $g$, $I_{xy}$ is larger than the sum of the environmental information $I_\mathrm{env}$ and the internal information $I_\mathrm{int}$, whereas it is smaller for $g<0$. (f) The sign of $g$ controls information interference, which is constructive if $g>0$, and destructive if $g<0$. In this panel, $D_\gamma = 10$. (g) In a slow bath $\tau_\eta \gg \tau$, constructive interference diverges for all $D_\gamma$ as $g \to 1$, whereas destructive interference remains finite at finite $D_\gamma$ when $g \to -1$.}
    \label{fig:harmonic_potentials}
\end{figure*}

\noindent We first focus on the analytically tractable scenario where the interactions between the particles are described by a harmonic potential, i.e.,
\begin{equation}
    F_x(x,y) = y, \quad F_y(x,y) = x \; .
\end{equation}
The stationary covariance of the system obeys the Lyapunov equation introduced in Eq.~\eqref{eqn:lyap}, where the interaction matrix now reads
\begin{equation*}
    \hat{A} = \begin{pmatrix}
        1/\tau & -g/\tau & -\gamma/\tau \\
        -g/\tau & 1/\tau & -\gamma/\tau \\
        0 & 0 & 1/\tau_\eta \\
    \end{pmatrix} \; .
\end{equation*}
Thus, for $g^2 < 1$, we can write the stationary solution for the internal degrees of freedom, $x$ and $y$, as the following Gaussian distribution,
\begin{equation*}
    p_{xy} = \mathcal{N}(0, \hat{\Sigma}^\mathrm{int}),
\end{equation*}
where the superscript denotes the presence of an internal coupling between $x$ and $y$. $\hat{\Sigma}^\mathrm{int}$ again corresponds to the submatrix of $\hat{\Sigma}$ associated with the degrees of freedom considered, as in the previous section. The model is depicted in Figure \ref{fig:harmonic_potentials}a.

For the sake of simplicity, from now on we take $D_x = D_y = 1$. We can immediately compute the mutual information $I_{xy}$ from the covariance matrix, as we did in  Eq.~\eqref{eqn:Ixy_env}. In this case, it reads as follows:
\begin{equation}
\label{eqn:Ixy_linear}
    I_{xy}= \frac{1}{2}\log\frac{\left[\alpha D_\gamma (g+1)-\zeta_\alpha(g)\right]^2}{(1 - g^2)\zeta_\alpha(g)\left[\zeta_\alpha(g) - 2 D_\gamma \alpha \right]}
\end{equation}
where $\zeta_\alpha(g) = \alpha (g-1)-1$. In particular, the internal mutual information can be obtained for $D_\gamma = 0$, i.e., when the active bath is not coupled to $x$ and $y$. Thus:
\begin{equation}
\label{eqn:Ixy_int_linear}
    I_{xy}^\mathrm{int} = \frac{1}{2}\log \frac{1}{1 - g^2} \; .
\end{equation}
$I_{xy}^\mathrm{int}$ diverges at the edge of the stability of the system, that is $g \to \pm 1$, as shown in previous works \cite{barzon2024maximal}, and, as expected, is symmetric under changes in the sign of $g$. Crucially, we notice that there exists a value of the timescale ratio $\alpha$ for which the mutual information has a non-trivial zero, namely
\begin{equation}
    \alpha_0 = -\frac{g}{D_\gamma + g(1 - g + D_\gamma)}
\end{equation}
which is positive if and only if $g < 0$. Thus, when the timescale of the active bath is exactly $\tau_\eta = \alpha_0 \tau$, the interference term balances exactly the environmental and internal information, canceling out any statistical dependency between the particles. In Figures \ref{fig:harmonic_potentials}b-c, we show the mutual information as a function of $D_\gamma$ and $\alpha$ for a positive and a negative value of the coupling $g$, showing the band of total destructive interference in the latter.

In general, we can explicitly write the information interference term as
\begin{equation}
    \label{eqn:xi_linear}
    \Xi_{xy} = \frac{1}{2}\log \frac{\lambda^{(1)}_\alpha(D_\gamma)\left[\alpha D_\gamma (g+1)-\zeta_\alpha(g)\right]^2}{\zeta_\alpha(g)\left[\zeta_\alpha(g) - 2 D_\gamma \alpha \right]\left[\lambda^{(1)}_\alpha(D_\gamma) + D_\gamma^2\alpha^2\right]}
\end{equation}
where $\lambda^{(1)}_\alpha(D_\gamma) = (\alpha(1 + D_\gamma) + 1)(1+\alpha)$ is a term depending solely on the environment, and $\zeta_\alpha(g)$ solely on interactions. In Figure \ref{fig:harmonic_potentials}d-e, we show how information interference may be either positive or negative, respectively increasing or reducing $I_{xy}$ with respect to the baseline $I_{xy}^\mathrm{int} + I_{xy}^\mathrm{env}$. Eq.~\eqref{eqn:xi_linear} has several interesting properties. In particular, it can be shown analytically that $\partial_{D_\gamma} |\Xi_{xy}| \ge 0$, so that the amount of information interference increases with the coupling to the active bath. This behavior is a signature of the fact that the activity of the environment is triggering the onset of information interference even if we are dealing with a linear scenario. Similarly, $|\Xi_{xy}|$ increases with $\alpha$, implying that it is larger the slower the environment. Notably,
\begin{equation}
    \lim_{\alpha \to 0} \Xi_{xy} = 0 \; .
\end{equation}
This result is not surprising. Indeed, directed interactions stemming from a fast degree of freedom - the active noise $\eta$, in this case - do not generate information, as shown in Eq.~\eqref{eqn:Ienv_limits} and more in general in \cite{nicoletti2024information}. As a consequence, a fast active environment only influences the dynamics of the particles, but not their information - thus, $I_{xy} = I_{xy}^\mathrm{int}$ when $\alpha \to 0$.

\begin{figure*}
    \centering
    \includegraphics[width=\textwidth]{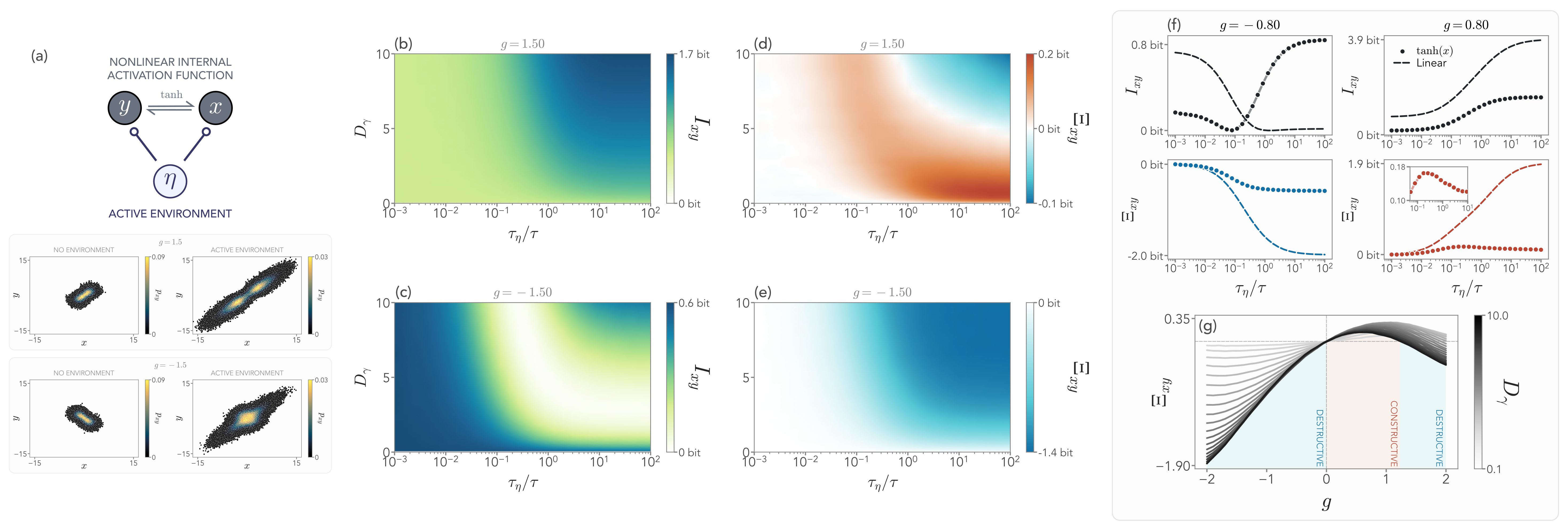}
    \caption{(a) Internal interactions between the particles are modeled with a nonlinear activation function. For large enough internal coupling $g$ and in the absence of the environment, the stationary probability of the system displays two peaks, either correlated ($g = 1.5$) or anticorrelated ($g = -1.5$). The active environment always tends to correlate the particles instead. (b-c) Similarly to the linear case, for $g = 1.5$ the mutual information between $x$ and $y$ increases both with $\alpha = \tau_\eta /\tau$ and with $D_\gamma = \gamma^2 D_\eta$. For $g = -1.5$, instead, $I_{xy}$ is non-monotonic in $\tau_\eta /\tau$. (d-e) Although information interference is always destructive for $g = -1.5$, the nonlinearity of internal interactions results in a switch between constructive and destructive interference for $g = 1.5$, depending on $\alpha$ and $D_\gamma$. (f) For small internal couplings $-1 < g < 1$ both $I_{xy}$ and $\Xi_{xy}$ generally display a qualitatively similar behavior with respect to the linear case. However, for $g < 0$, the timescale at which total destructive interference is achieved is markedly different. Furthermore, for positive $g$, constructive interference displays a peak at intermediate timescales. In this panel, $D_\gamma = 10$. (g) The switch between destructive and constructive interference takes place at large enough $g$ and depends on the bath coupling $D_\gamma$. In this panel, $\alpha = 10^2$.}
    \label{fig:tanh}
\end{figure*}

In Figure \ref{fig:harmonic_potentials}f, we show a striking dependence of the sign of $\Xi_{xy}$ on that of the internal coupling $g$. In particular, for $g>0$ we can only find constructive information interference, while for $g<0$ the system may only display destructive interference. This spontaneous breaking of the symmetry under changes of the sign of $g$ is in stark contrast with the phenomenology associated with the absence of an active environment, Eq.~\eqref{eqn:Ixy_int_linear}. Heuristically, one can understand this result in terms of the different roles that $g$ and $\gamma$ play in the system. While the active noise tends to create positive correlations between the particles, internal interactions may both correlate or anticorrelate them, depending on the sign of $g$. Thus, for $g<0$ the two effects are competing, masking each other and leading to destructive information interference. On the other hand, for $g>0$, the active environment acts synergistically with internal interactions, generating more information than the sum of the two. Notably, as the system approaches the edge of stability, we find that
\begin{gather}
        \Xi_{xy}(g \to -1) = \frac{1}{2}\log\frac{\lambda^{(1)}_\alpha(D_\gamma)(1 + 2 \alpha)}{[1 + 2 \alpha(1 +  D_\gamma)](\lambda^{(1)}_\alpha(D_\gamma)+D_\gamma^2\alpha^2)} \nonumber \\
        \Xi_{xy}(g \to 1) = \frac{1}{2}\log\frac{\lambda^{(1)}_\alpha(D_\gamma)(1 + 2 \alpha D_\gamma)}{\lambda^{(1)}_\alpha(D_\gamma)+D_\gamma^2\alpha^2} \; .
\end{gather}
In Figure \ref{fig:harmonic_potentials}g, we plot these two expressions in the limit of a slow environment $\alpha \to \infty$. As the system approaches the edge of stability for $g \to 1$, both the mutual information and constructive information interference diverge, as expected. On the other hand, when instability is approached from $g \to -1$, destructive information interference remains finite - since it is lower bounded, as discussed above - and diverges only for a large coupling to the active bath, $D_\gamma \to \infty$.

\section{Information interference in nonlinear force fields}
\noindent We now switch to a case in which the interactions between the two particles are neither linear nor conservative, so that mutual information captures nonlinear dependencies beyond simple correlations. We take the force fields to be
\begin{equation}
    F_x(x,y) = \tanh(y), \quad F_y(x,y) = \tanh(x) \; ,
\end{equation}
where $\tanh$ describes a generic activation function that has been employed in various modeling scenarios \cite{sharma2017activation,christodoulou2022regimes,kunc2021transformative}. Although the system does not admit an analytical solution anymore, we can study it numerically. In particular, to estimate the mutual information, we first simulate the trajectories from the Langevin equations, Eq.~\eqref{eqn:model}, and then use a k-nearest neighbor estimator to obtain $I_{xy}$ \cite{kraskov2004estimating}. If not otherwise specified, trajectories are simulated using a standard Euler-Mayorama algorithm for $10^6$ steps with a step size of $dt = 10^{-3}$, and the mutual information for a set of parameters is obtained by sampling $10^3$ trajectories at large time intervals to avoid the bias induced by the autocorrelation of subsequent time points. In Figure \ref{fig:tanh}a, we sketch the system and plot the stationary probability distribution $p_{xy}$ for different coupling values $g$. In the absence of the active noise, for large enough $|g|$ the system has two fixed points. If $g$ is positive, both the fixed points of $x$ and $y$ share the same sign - heuristically, they are correlated - whereas they have opposite signs at negative $g$ - i.e., the particles are anticorrelated. When the active noise is present, the probability distribution is altered significantly, as the shared environment tends to concentrate the probability along $x = y$.

In Figure \ref{fig:tanh}b-c, we plot $I_{xy}$ as a function of $D_\gamma$ and $\alpha$ for $g= \pm 1.5$. We seemingly find a qualitative behavior that is similar to the linear case, with a band of total destructive interference for negative $g$, and a mutual information monotonically increasing with $\alpha$ for positive $g$. However, the picture becomes markedly different when considering the information interference term, plotted in Figure \ref{fig:tanh}d-e. Although, as expected, destructive interference occurs at $g = -1.5$, for $g = 1.5$ and high enough $D_\gamma$ we find a switch between constructive and destructive interference depending on the timescales of the system. This surprising result signals that, although both the total mutual information $I_{xy}$ and the environmental one $I_{xy}^\mathrm{env}$ - that does not depend on interactions by construction - increase with $\alpha$, $I_{xy}^\mathrm{env}$ increases more rapidly, eventually pushing the interference term to become negative.

In Figure \ref{fig:tanh}f-g, we show that this phenomenon only appears at large enough internal coupling. When $g \in (-1, 1)$, both $I_{xy}$ and $\Xi_{xy}$ display similar behaviors between the linear and the nonlinear case, with a few notable differences (Figure \ref{fig:tanh}f). In particular, at negative $g$ we find that the presence of nonlinear interactions changes the timescale ratio $\alpha_0$ at which total destructive interference occurs. Furthermore, in the nonlinear case, destructive interference is severely reduced in the limit of a slow environment. Similarly, when $g>0$ the nonlinear constructive interference term displays a peak at intermediate values of $\alpha$, and it is typically smaller with respect to the linear case. 

When $g$ becomes larger, nonlinear effects start being dominant. The eventual switch from constructive to destructive interference depends on $D_\gamma$ (see Figure \ref{fig:tanh}g), as the interplay between internal interactions and environmental changes becomes more complex due to the saturating effects of the activation function. Indeed, even in the absence of the active noise, as $|g| \to \infty$ both $x$ and $y$ increase in absolute value, leading the $\tanh$ to quickly saturate. Thus, the nonlinear interactions reduce to two constant drifts, effectively leaving the particles decoupled, with $I_{xy}^\mathrm{int} = 0$. Although not shown in Figure \ref{fig:tanh} for simplicity, this leads to complex switches between constructive and destructive interference.

\section{Information interference in hierarchical systems}
\noindent Finally, we study the slightly different setting of a hierarchical system, where $y$ interacts with $x$, but not vice-versa. Furthermore, we assume that $y$ may be weakly coupled to the bath, so that
\begin{equation}
\label{eqn:model_hier}
    \begin{gathered}
        \tau \dot{x}(t) = -x(t) + \sqrt{2 \tau} \xi_x(t) + \gamma \eta(t) \\
        \tau \dot{y}(t) = -y(t) + g F_y(x) + \sqrt{2 \tau} \xi_y(t) + \epsilon\gamma \eta(t)
    \end{gathered}
\end{equation}
where $\epsilon \in [0,1]$. If $\epsilon = 0$, the system is fully hierarchical, as the interactions follow $\eta \to x \to y$, a setting used to study transduction mechanisms \cite{nicoletti2024tuning}, where the environment is accessible only to $x$. The more general scenario of Eq.~\eqref{eqn:model_hier} mimics the fact that the environment might be partially hidden to one degree of freedom - $y$ in this case - with $\epsilon$ quantifying the degree of accessibility. The environmental information can be obtained by setting to zero internal interactions, as before, and it reads:
\begin{equation}
\label{eqn:Ienv_hier}
    I_{xy}^\mathrm{env} = \frac{1}{2}\log\frac{[1 + \alpha(D_\gamma + 1)][1 + \alpha(D_\gamma \epsilon^2 + 1)]}{(1 + \alpha)[1 + \alpha + \alpha D_\gamma(1 + \epsilon^2)]} \; .
\end{equation}
As for the previous sections, we study the cases of $F_y(x) = x$ and $F_y(x) = \tanh(x)$ and focus on how $\epsilon$ modulates information and interference. 

\begin{figure}
    \centering
    \includegraphics[width=\columnwidth]{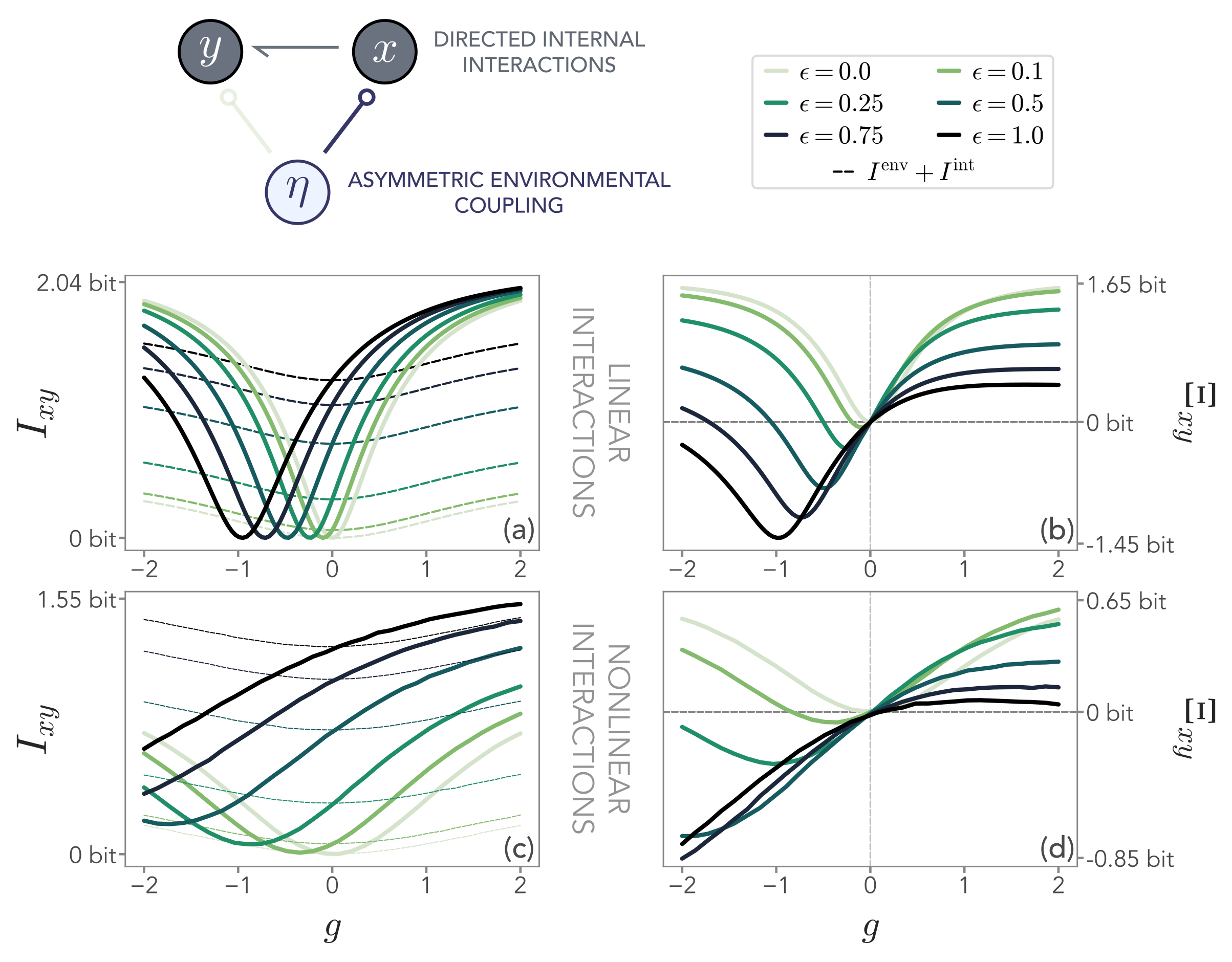}
    \caption{(a-b) Mutual information and information interference in a hierarchical system where the two particles have different couplings with the active noise (Eq.~\eqref{eqn:model_hier}) and only $y$ interacts with $x$. For linear interactions, increasing the coupling of $y$ with the environment, $\epsilon$, shifts the minimum of $I_{xy}$ to negative $g$, and destructive information interference appears. (c-d) When the interaction from $x$ to $y$ is the nonlinear activation function $\tanh(x)$, we find a similar phenomenology. Notably, for the values of $g$ explored here, no total destructive information appears anymore, contrary to the linear case. In these plots, $D_\gamma = 10$ and $\alpha = 10^2$.}
    \label{fig:hierarchical}
\end{figure}

When $\epsilon = 0$, no shared coupling with the active environment is present and $y$ evolves independently. As a consequence, we immediately have that $I_{xy}^\mathrm{env} = 0$ from Eq.~\eqref{eqn:Ienv_hier}. In the linear case, the interference term reads
\begin{equation}
    \Xi_{xy}(\epsilon = 0) = \frac{1}{2}\log\frac{4+g^2}{2+g^2}\frac{\lambda_\alpha\left[2(1+\alpha)^2 + g^2 \lambda_\alpha^{(2)}\right]}{4 \lambda_\alpha (1+\alpha)^2 + g^2 \lambda_\alpha^{(2)}\lambda_\alpha^{(3)}}
\end{equation}
where we introduced the environmental functions $\lambda_\alpha^{(2)} = 1 + \alpha (2 + D_\gamma + \alpha)$ and $\lambda_\alpha^{(3)} = 1 + \alpha(2 + D_\gamma + \alpha (1 + 2D_\gamma))$. We have that $\Xi_{xy}(\epsilon = 0) > 0$, so that information interference in this fully hierarchical system can only be constructive. This is perhaps unsurprising since the active environment is not shared anymore and thus it cannot mask the information generated by the interactions, thereby avoiding the regime of destructive interference. As we show in Figure \ref{fig:hierarchical}a-b, when we turn on the coupling between the active environment and $y$, a region of destructive information interference appears and it broadens as $\epsilon$ increases. Notably, in contrast with the equilibrium harmonic potential studied before, here the system is stable for all values of $g$. We find that there exists a critical value of the coupling $g < 0$ for which information interference switches once more from destructive to constructive, depending on $\epsilon$. At these critical couplings, $\Xi_{xy} = 0$, and the mutual information between the particles becomes exactly the sum of the internal information and the environmental one, with the two contributions being fully disentangled \cite{nicoletti2021mutual}. Finally, in Figure \ref{fig:hierarchical}c-d we show that the mutual information displays a qualitatively similar behavior in the presence of a nonlinear activation function, $F_y = \tanh(x)$. Quantitatively, at the same value of $\epsilon$, the region of couplings at which destructive information interference is present is significantly broadened, while the overall amount of information interference tends to be lower in the parameter regimes we explored. Indeed, for $\epsilon \neq 0$, total destructive interference is not achieved anymore for negative $g$ (see Figure \ref{fig:hierarchical}c). Thus, the saturating effect of the nonlinearity fosters destructive interference while limiting the overall magnitude of $\Xi$.

\section{Discussions}
\noindent In this work, we studied how the coupling to a shared active environment modulates the mutual information, i.e., the statistical dependencies of two interacting Brownian particles $x$ and $y$. The active environment $\eta$ is modeled by a colored noise with autocorrelation decaying exponentially with a characteristic timescale $\tau_\eta$. We exploited the decomposition first proposed in \cite{nicoletti2022mutual} to identify three different contributions to the mutual information between $x$ and $y$ (Eq.~\eqref{eqn:interference}). The first term is the internal information $I_{xy}$, i.e., the information that is solely due to internal interactions. This quantity can be computed exactly in the linear case (Eq.~\eqref{eqn:Ixy_int_linear}) and numerically for more general scenarios. The second term is the information induced by the shared environmental coupling alone, which can be obtained analytically (Eq.~\eqref{eqn:Ixy_env}), showing that the active bath may couple the particles even if they are not interacting directly. The last term is an interference term that models the interplay occurring when both internal interactions and the active environment are present simultaneously. In particular, this interference can be either constructive - with the dependency between particles amplified by the presence of both interactions and environment at once - or destructive, up to the point of total destructive interference, where the environmental effects cancel exactly the internal ones, leaving the particles effectively independent.

We have shown that, depending on the coupling to the active bath and on the relative timescales between the internal dynamics and the environmental one, both constructive and destructive interference may occur. In the presence of linear interactions, interference is fully determined by the sign of the internal coupling, introducing a novel asymmetry that would not be present without the active noise term. The interplay between timescales and couplings becomes more complex when internal interactions are expressed through a nonlinear activation function, with multiple non-trivial switches between destructive and constructive interference. Finally, we focus on the case of an asymmetric coupling to the bath mimicking the partial accessibility of the active environment, showing that the dependencies induced by a shared bath are essential to achieve destructive interference both in the linear and nonlinear case. Overall, our work demonstrates that introducing the concept of information interference enables a clear quantification of the structure of statistical dependencies in a physical system, breaking them down into interpretable contributions.

We remark that, contrarily to previous studies that focused on environments that act on, e.g., the diffusion coefficient to model diffusion in random media, temperature fluctuations, or different neural activity regimes \cite{chechkin2017brownian, wang2020unexpected, nicoletti2021mutual, mariani2022disentangling, honey2017switching, nicoletti2020scaling}, the case of an active environment considered here is fundamentally different. The active noise term in Eq.~\eqref{eqn:model} is essentially an interaction term following an additional Ornstein-Uhelnbeck process. As such, it formally acts in the same way as internal interactions. Therefore, in the scenario studied here, the distinction between internal interactions and environmental processes depends on the way we choose to marginalize the joint probability distribution, and disentangling the two contributions is in general not possible unless for particular parameter choices for which $\Xi = 0$. 

In future works, it will be interesting to study how information and information interference behave depending on how the system is partitioned, allowing for a deeper understanding of different sources of dependencies in complex systems composed of multiple interacting parts. Our results may indeed be applied to several fields: from ecology, studying the interplay between landscape changes, species interactions, and dispersal \cite{hastings2010timescales, poisot2015beyond, nicoletti2023emergent, padmanabha2024spatially}, to neuroscience, where neural activity is influenced by external stimuli or extrinsic activity stemming from interactions with other brain regions that are cannot be directly measured \cite{ferrari2018separating, mariani2022disentangling}, and it is shaped by a network of interactions \cite{sporns2011human, mivsic2015cooperative, barzon2022criticality} that may evolve due to plasticity \cite{stepanyants2002geometry, stampanoni2019synaptic}. In general, many natural systems can be immediately partitioned into separate, interpretable components that modify their overall behavior, due to their intrinsic multiscale structure \cite{schaffer2021mapping, boccaletti2014structure, bianconi2018multilayer, aleta2019multilayer, artime2022multilayer}. 

Moreover, this study lays the foundation for the analysis of how nonequilibrium conditions shape functional relationships in complex biological systems. Indeed, the link between information and thermodynamics dates back to seminal works in the last decade \cite{parrondo2015thermodynamics,sagawa2012thermodynamics,horowitz2014thermodynamics}. However, how dissipative phenomena constrain the emergence of interactions that effectively spread information throughout the system is still unclear and underexplored. This work proposes a general framework, rooted in the concept of information interference, that may be informative to study this problem. The interplay between information interference and entropy production, or any other signature of nonequilibrium conditions, will be an interesting subject of future investigations.

In summary, our work paves the way for a general understanding of how multiple processes at possibly different timescales may both compete or cooperate to define functional relationships in a physical system. Even a simple coupling with an active bath leads to a rich structure of statistical dependencies determined by the interplay between the timescales at hand, the coupling between the system and the bath, and the shape of the internal interactions in the system. This paradigmatic example underscores that functional relationships in many complex systems are often context-dependent, i.e., that the environmental context plays a fundamental role in determining information.

\begin{acknowledgments}
\noindent G.N. acknowledges funding provided by the Swiss National Science Foundation through its Grant CRSII5\_186422. The authors acknowledge the support of the Munich Institute for Astro-, Particle and BioPhysics (MIAPbP), funded by the Deutsche Forschungsgemeinschaft (DFG, German Research Foundation) under Germany´s Excellence Strategy – EXC-2094 – 390783311, where this work was first conceived during the MOLINFO workshop.
\end{acknowledgments}


\begin{thebibliography}{66}%
\makeatletter
\providecommand \@ifxundefined [1]{%
 \@ifx{#1\undefined}
}%
\providecommand \@ifnum [1]{%
 \ifnum #1\expandafter \@firstoftwo
 \else \expandafter \@secondoftwo
 \fi
}%
\providecommand \@ifx [1]{%
 \ifx #1\expandafter \@firstoftwo
 \else \expandafter \@secondoftwo
 \fi
}%
\providecommand \natexlab [1]{#1}%
\providecommand \enquote  [1]{``#1''}%
\providecommand \bibnamefont  [1]{#1}%
\providecommand \bibfnamefont [1]{#1}%
\providecommand \citenamefont [1]{#1}%
\providecommand \href@noop [0]{\@secondoftwo}%
\providecommand \href [0]{\begingroup \@sanitize@url \@href}%
\providecommand \@href[1]{\@@startlink{#1}\@@href}%
\providecommand \@@href[1]{\endgroup#1\@@endlink}%
\providecommand \@sanitize@url [0]{\catcode `\\12\catcode `\$12\catcode `\&12\catcode `\#12\catcode `\^12\catcode `\_12\catcode `\%12\relax}%
\providecommand \@@startlink[1]{}%
\providecommand \@@endlink[0]{}%
\providecommand \url  [0]{\begingroup\@sanitize@url \@url }%
\providecommand \@url [1]{\endgroup\@href {#1}{\urlprefix }}%
\providecommand \urlprefix  [0]{URL }%
\providecommand \Eprint [0]{\href }%
\providecommand \doibase [0]{https://doi.org/}%
\providecommand \selectlanguage [0]{\@gobble}%
\providecommand \bibinfo  [0]{\@secondoftwo}%
\providecommand \bibfield  [0]{\@secondoftwo}%
\providecommand \translation [1]{[#1]}%
\providecommand \BibitemOpen [0]{}%
\providecommand \bibitemStop [0]{}%
\providecommand \bibitemNoStop [0]{.\EOS\space}%
\providecommand \EOS [0]{\spacefactor3000\relax}%
\providecommand \BibitemShut  [1]{\csname bibitem#1\endcsname}%
\let\auto@bib@innerbib\@empty
\bibitem [{\citenamefont {Hilfinger}\ and\ \citenamefont {Paulsson}(2011)}]{Hilfinger2011}%
  \BibitemOpen
  \bibfield  {author} {\bibinfo {author} {\bibfnamefont {A.}~\bibnamefont {Hilfinger}}\ and\ \bibinfo {author} {\bibfnamefont {J.}~\bibnamefont {Paulsson}},\ }\bibfield  {title} {\bibinfo {title} {Separating intrinsic from extrinsic fluctuations in dynamic biological systems},\ }\href {https://doi.org/10.1073/pnas.1018832108} {\bibfield  {journal} {\bibinfo  {journal} {Proceedings of the National Academy of Sciences}\ }\textbf {\bibinfo {volume} {108}},\ \bibinfo {pages} {12167} (\bibinfo {year} {2011})}\BibitemShut {NoStop}%
\bibitem [{\citenamefont {Tsimring}(2014)}]{Tsimring2014}%
  \BibitemOpen
  \bibfield  {author} {\bibinfo {author} {\bibfnamefont {L.~S.}\ \bibnamefont {Tsimring}},\ }\bibfield  {title} {\bibinfo {title} {Noise in biology},\ }\href {https://doi.org/10.1088/0034-4885/77/2/026601} {\bibfield  {journal} {\bibinfo  {journal} {Reports on Progress in Physics}\ }\textbf {\bibinfo {volume} {77}},\ \bibinfo {pages} {026601} (\bibinfo {year} {2014})}\BibitemShut {NoStop}%
\bibitem [{\citenamefont {Zhu}\ and\ \citenamefont {Yin}(2009)}]{Zhu2009}%
  \BibitemOpen
  \bibfield  {author} {\bibinfo {author} {\bibfnamefont {C.}~\bibnamefont {Zhu}}\ and\ \bibinfo {author} {\bibfnamefont {G.}~\bibnamefont {Yin}},\ }\bibfield  {title} {\bibinfo {title} {On competitive lotka–volterra model in random environments},\ }\href {https://doi.org/https://doi.org/10.1016/j.jmaa.2009.03.066} {\bibfield  {journal} {\bibinfo  {journal} {Journal of Mathematical Analysis and Applications}\ }\textbf {\bibinfo {volume} {357}},\ \bibinfo {pages} {154} (\bibinfo {year} {2009})}\BibitemShut {NoStop}%
\bibitem [{\citenamefont {Wienand}\ \emph {et~al.}(2017)\citenamefont {Wienand}, \citenamefont {Frey},\ and\ \citenamefont {Mobilia}}]{wienand2017fluctuating}%
  \BibitemOpen
  \bibfield  {author} {\bibinfo {author} {\bibfnamefont {K.}~\bibnamefont {Wienand}}, \bibinfo {author} {\bibfnamefont {E.}~\bibnamefont {Frey}},\ and\ \bibinfo {author} {\bibfnamefont {M.}~\bibnamefont {Mobilia}},\ }\bibfield  {title} {\bibinfo {title} {Evolution of a fluctuating population in a randomly switching environment},\ }\href@noop {} {\bibfield  {journal} {\bibinfo  {journal} {Physical Review Letters}\ }\textbf {\bibinfo {volume} {119}},\ \bibinfo {pages} {158301} (\bibinfo {year} {2017})}\BibitemShut {NoStop}%
\bibitem [{\citenamefont {Wienand}\ \emph {et~al.}(2018)\citenamefont {Wienand}, \citenamefont {Frey},\ and\ \citenamefont {Mobilia}}]{wienand2018eco}%
  \BibitemOpen
  \bibfield  {author} {\bibinfo {author} {\bibfnamefont {K.}~\bibnamefont {Wienand}}, \bibinfo {author} {\bibfnamefont {E.}~\bibnamefont {Frey}},\ and\ \bibinfo {author} {\bibfnamefont {M.}~\bibnamefont {Mobilia}},\ }\bibfield  {title} {\bibinfo {title} {Eco-evolutionary dynamics of a population with randomly switching carrying capacity},\ }\href@noop {} {\bibfield  {journal} {\bibinfo  {journal} {Journal of The Royal Society Interface}\ }\textbf {\bibinfo {volume} {15}},\ \bibinfo {pages} {20180343} (\bibinfo {year} {2018})}\BibitemShut {NoStop}%
\bibitem [{\citenamefont {Padmanabha}\ \emph {et~al.}(2024)\citenamefont {Padmanabha}, \citenamefont {Nicoletti}, \citenamefont {Bernardi}, \citenamefont {Suweis}, \citenamefont {Azaele}, \citenamefont {Rinaldo},\ and\ \citenamefont {Maritan}}]{padmanabha2024spatially}%
  \BibitemOpen
  \bibfield  {author} {\bibinfo {author} {\bibfnamefont {P.}~\bibnamefont {Padmanabha}}, \bibinfo {author} {\bibfnamefont {G.}~\bibnamefont {Nicoletti}}, \bibinfo {author} {\bibfnamefont {D.}~\bibnamefont {Bernardi}}, \bibinfo {author} {\bibfnamefont {S.}~\bibnamefont {Suweis}}, \bibinfo {author} {\bibfnamefont {S.}~\bibnamefont {Azaele}}, \bibinfo {author} {\bibfnamefont {A.}~\bibnamefont {Rinaldo}},\ and\ \bibinfo {author} {\bibfnamefont {A.}~\bibnamefont {Maritan}},\ }\bibfield  {title} {\bibinfo {title} {Spatially disordered environments stabilize competitive metacommunities},\ }\href@noop {} {\bibfield  {journal} {\bibinfo  {journal} {arXiv preprint arXiv:2404.09908}\ } (\bibinfo {year} {2024})}\BibitemShut {NoStop}%
\bibitem [{\citenamefont {Swain}\ \emph {et~al.}(2002)\citenamefont {Swain}, \citenamefont {Elowitz},\ and\ \citenamefont {Siggia}}]{swain2002extrinsic}%
  \BibitemOpen
  \bibfield  {author} {\bibinfo {author} {\bibfnamefont {P.~S.}\ \bibnamefont {Swain}}, \bibinfo {author} {\bibfnamefont {M.~B.}\ \bibnamefont {Elowitz}},\ and\ \bibinfo {author} {\bibfnamefont {E.~D.}\ \bibnamefont {Siggia}},\ }\bibfield  {title} {\bibinfo {title} {Intrinsic and extrinsic contributions to stochasticity in gene expression},\ }\href@noop {} {\bibfield  {journal} {\bibinfo  {journal} {Proceedings of the National Academy of Sciences}\ }\textbf {\bibinfo {volume} {99}},\ \bibinfo {pages} {12795} (\bibinfo {year} {2002})}\BibitemShut {NoStop}%
\bibitem [{\citenamefont {Thomas}\ \emph {et~al.}(2014)\citenamefont {Thomas}, \citenamefont {Popovi{\'c}},\ and\ \citenamefont {Grima}}]{thomas2014phenotypic}%
  \BibitemOpen
  \bibfield  {author} {\bibinfo {author} {\bibfnamefont {P.}~\bibnamefont {Thomas}}, \bibinfo {author} {\bibfnamefont {N.}~\bibnamefont {Popovi{\'c}}},\ and\ \bibinfo {author} {\bibfnamefont {R.}~\bibnamefont {Grima}},\ }\bibfield  {title} {\bibinfo {title} {Phenotypic switching in gene regulatory networks},\ }\href@noop {} {\bibfield  {journal} {\bibinfo  {journal} {Proceedings of the National Academy of Sciences}\ }\textbf {\bibinfo {volume} {111}},\ \bibinfo {pages} {6994} (\bibinfo {year} {2014})}\BibitemShut {NoStop}%
\bibitem [{\citenamefont {Bowsher}\ and\ \citenamefont {Swain}(2012)}]{bowsher2012biochemical}%
  \BibitemOpen
  \bibfield  {author} {\bibinfo {author} {\bibfnamefont {C.~G.}\ \bibnamefont {Bowsher}}\ and\ \bibinfo {author} {\bibfnamefont {P.~S.}\ \bibnamefont {Swain}},\ }\bibfield  {title} {\bibinfo {title} {Identifying sources of variation and the flow of information in biochemical networks},\ }\href@noop {} {\bibfield  {journal} {\bibinfo  {journal} {Proceedings of the National Academy of Sciences}\ }\textbf {\bibinfo {volume} {109}},\ \bibinfo {pages} {E1320} (\bibinfo {year} {2012})}\BibitemShut {NoStop}%
\bibitem [{\citenamefont {Dass}\ \emph {et~al.}(2021)\citenamefont {Dass}, \citenamefont {Georgelin}, \citenamefont {Westall}, \citenamefont {Foucher}, \citenamefont {De~Los~Rios}, \citenamefont {Busiello}, \citenamefont {Liang},\ and\ \citenamefont {Piazza}}]{dass2021equilibrium}%
  \BibitemOpen
  \bibfield  {author} {\bibinfo {author} {\bibfnamefont {A.~V.}\ \bibnamefont {Dass}}, \bibinfo {author} {\bibfnamefont {T.}~\bibnamefont {Georgelin}}, \bibinfo {author} {\bibfnamefont {F.}~\bibnamefont {Westall}}, \bibinfo {author} {\bibfnamefont {F.}~\bibnamefont {Foucher}}, \bibinfo {author} {\bibfnamefont {P.}~\bibnamefont {De~Los~Rios}}, \bibinfo {author} {\bibfnamefont {D.~M.}\ \bibnamefont {Busiello}}, \bibinfo {author} {\bibfnamefont {S.}~\bibnamefont {Liang}},\ and\ \bibinfo {author} {\bibfnamefont {F.}~\bibnamefont {Piazza}},\ }\bibfield  {title} {\bibinfo {title} {Equilibrium and non-equilibrium furanose selection in the ribose isomerisation network},\ }\href@noop {} {\bibfield  {journal} {\bibinfo  {journal} {Nature communications}\ }\textbf {\bibinfo {volume} {12}},\ \bibinfo {pages} {2749} (\bibinfo {year} {2021})}\BibitemShut {NoStop}%
\bibitem [{\citenamefont {Dorsaz}\ \emph {et~al.}(2010)\citenamefont {Dorsaz}, \citenamefont {De~Michele}, \citenamefont {Piazza}, \citenamefont {De~Los~Rios},\ and\ \citenamefont {Foffi}}]{dorsaz2010diffusion}%
  \BibitemOpen
  \bibfield  {author} {\bibinfo {author} {\bibfnamefont {N.}~\bibnamefont {Dorsaz}}, \bibinfo {author} {\bibfnamefont {C.}~\bibnamefont {De~Michele}}, \bibinfo {author} {\bibfnamefont {F.}~\bibnamefont {Piazza}}, \bibinfo {author} {\bibfnamefont {P.}~\bibnamefont {De~Los~Rios}},\ and\ \bibinfo {author} {\bibfnamefont {G.}~\bibnamefont {Foffi}},\ }\bibfield  {title} {\bibinfo {title} {Diffusion-limited reactions in crowded environments},\ }\href@noop {} {\bibfield  {journal} {\bibinfo  {journal} {Physical Review Letters}\ }\textbf {\bibinfo {volume} {105}},\ \bibinfo {pages} {120601} (\bibinfo {year} {2010})}\BibitemShut {NoStop}%
\bibitem [{\citenamefont {Galanti}\ \emph {et~al.}(2014)\citenamefont {Galanti}, \citenamefont {Fanelli}, \citenamefont {Maritan},\ and\ \citenamefont {Piazza}}]{galanti2014diffusion}%
  \BibitemOpen
  \bibfield  {author} {\bibinfo {author} {\bibfnamefont {M.}~\bibnamefont {Galanti}}, \bibinfo {author} {\bibfnamefont {D.}~\bibnamefont {Fanelli}}, \bibinfo {author} {\bibfnamefont {A.}~\bibnamefont {Maritan}},\ and\ \bibinfo {author} {\bibfnamefont {F.}~\bibnamefont {Piazza}},\ }\bibfield  {title} {\bibinfo {title} {Diffusion of tagged particles in a crowded medium},\ }\href@noop {} {\bibfield  {journal} {\bibinfo  {journal} {EPL (Europhysics Letters)}\ }\textbf {\bibinfo {volume} {107}},\ \bibinfo {pages} {20006} (\bibinfo {year} {2014})}\BibitemShut {NoStop}%
\bibitem [{\citenamefont {Chechkin}\ \emph {et~al.}(2017)\citenamefont {Chechkin}, \citenamefont {Seno}, \citenamefont {Metzler},\ and\ \citenamefont {Sokolov}}]{chechkin2017brownian}%
  \BibitemOpen
  \bibfield  {author} {\bibinfo {author} {\bibfnamefont {A.~V.}\ \bibnamefont {Chechkin}}, \bibinfo {author} {\bibfnamefont {F.}~\bibnamefont {Seno}}, \bibinfo {author} {\bibfnamefont {R.}~\bibnamefont {Metzler}},\ and\ \bibinfo {author} {\bibfnamefont {I.~M.}\ \bibnamefont {Sokolov}},\ }\bibfield  {title} {\bibinfo {title} {Brownian yet non-gaussian diffusion: from superstatistics to subordination of diffusing diffusivities},\ }\href@noop {} {\bibfield  {journal} {\bibinfo  {journal} {Physical Review X}\ }\textbf {\bibinfo {volume} {7}},\ \bibinfo {pages} {021002} (\bibinfo {year} {2017})}\BibitemShut {NoStop}%
\bibitem [{\citenamefont {Piazza}(2008)}]{piazza2008thermophoresis}%
  \BibitemOpen
  \bibfield  {author} {\bibinfo {author} {\bibfnamefont {R.}~\bibnamefont {Piazza}},\ }\bibfield  {title} {\bibinfo {title} {Thermophoresis: moving particles with thermal gradients},\ }\href@noop {} {\bibfield  {journal} {\bibinfo  {journal} {Soft Matter}\ }\textbf {\bibinfo {volume} {4}},\ \bibinfo {pages} {1740} (\bibinfo {year} {2008})}\BibitemShut {NoStop}%
\bibitem [{\citenamefont {Liang}\ \emph {et~al.}(2022)\citenamefont {Liang}, \citenamefont {Busiello},\ and\ \citenamefont {De~Los~Rios}}]{liang2022emergent}%
  \BibitemOpen
  \bibfield  {author} {\bibinfo {author} {\bibfnamefont {S.}~\bibnamefont {Liang}}, \bibinfo {author} {\bibfnamefont {D.~M.}\ \bibnamefont {Busiello}},\ and\ \bibinfo {author} {\bibfnamefont {P.}~\bibnamefont {De~Los~Rios}},\ }\bibfield  {title} {\bibinfo {title} {Emergent thermophoretic behavior in chemical reaction systems},\ }\href@noop {} {\bibfield  {journal} {\bibinfo  {journal} {New Journal of Physics}\ }\textbf {\bibinfo {volume} {24}},\ \bibinfo {pages} {123006} (\bibinfo {year} {2022})}\BibitemShut {NoStop}%
\bibitem [{\citenamefont {Busiello}\ \emph {et~al.}(2021)\citenamefont {Busiello}, \citenamefont {Liang}, \citenamefont {Piazza},\ and\ \citenamefont {De~Los~Rios}}]{busiello2021dissipation}%
  \BibitemOpen
  \bibfield  {author} {\bibinfo {author} {\bibfnamefont {D.~M.}\ \bibnamefont {Busiello}}, \bibinfo {author} {\bibfnamefont {S.}~\bibnamefont {Liang}}, \bibinfo {author} {\bibfnamefont {F.}~\bibnamefont {Piazza}},\ and\ \bibinfo {author} {\bibfnamefont {P.}~\bibnamefont {De~Los~Rios}},\ }\bibfield  {title} {\bibinfo {title} {Dissipation-driven selection of states in non-equilibrium chemical networks},\ }\href@noop {} {\bibfield  {journal} {\bibinfo  {journal} {Communications Chemistry}\ }\textbf {\bibinfo {volume} {4}},\ \bibinfo {pages} {16} (\bibinfo {year} {2021})}\BibitemShut {NoStop}%
\bibitem [{\citenamefont {Berton}\ \emph {et~al.}(2020)\citenamefont {Berton}, \citenamefont {Busiello}, \citenamefont {Zamuner}, \citenamefont {Solari}, \citenamefont {Scopelliti}, \citenamefont {Fadaei-Tirani}, \citenamefont {Severin},\ and\ \citenamefont {Pezzato}}]{berton2020thermodynamics}%
  \BibitemOpen
  \bibfield  {author} {\bibinfo {author} {\bibfnamefont {C.}~\bibnamefont {Berton}}, \bibinfo {author} {\bibfnamefont {D.~M.}\ \bibnamefont {Busiello}}, \bibinfo {author} {\bibfnamefont {S.}~\bibnamefont {Zamuner}}, \bibinfo {author} {\bibfnamefont {E.}~\bibnamefont {Solari}}, \bibinfo {author} {\bibfnamefont {R.}~\bibnamefont {Scopelliti}}, \bibinfo {author} {\bibfnamefont {F.}~\bibnamefont {Fadaei-Tirani}}, \bibinfo {author} {\bibfnamefont {K.}~\bibnamefont {Severin}},\ and\ \bibinfo {author} {\bibfnamefont {C.}~\bibnamefont {Pezzato}},\ }\bibfield  {title} {\bibinfo {title} {Thermodynamics and kinetics of protonated merocyanine photoacids in water},\ }\href@noop {} {\bibfield  {journal} {\bibinfo  {journal} {Chemical Science}\ }\textbf {\bibinfo {volume} {11}},\ \bibinfo {pages} {8457} (\bibinfo {year} {2020})}\BibitemShut {NoStop}%
\bibitem [{\citenamefont {Liang}\ \emph {et~al.}(2024)\citenamefont {Liang}, \citenamefont {De~Los~Rios},\ and\ \citenamefont {Busiello}}]{liang2024thermodynamic}%
  \BibitemOpen
  \bibfield  {author} {\bibinfo {author} {\bibfnamefont {S.}~\bibnamefont {Liang}}, \bibinfo {author} {\bibfnamefont {P.}~\bibnamefont {De~Los~Rios}},\ and\ \bibinfo {author} {\bibfnamefont {D.~M.}\ \bibnamefont {Busiello}},\ }\bibfield  {title} {\bibinfo {title} {Thermodynamic bounds on symmetry breaking in linear and catalytic biochemical systems},\ }\href@noop {} {\bibfield  {journal} {\bibinfo  {journal} {Physical Review Letters}\ }\textbf {\bibinfo {volume} {132}},\ \bibinfo {pages} {228402} (\bibinfo {year} {2024})}\BibitemShut {NoStop}%
\bibitem [{\citenamefont {Falasco}\ \emph {et~al.}(2018)\citenamefont {Falasco}, \citenamefont {Rao},\ and\ \citenamefont {Esposito}}]{falasco2018turing}%
  \BibitemOpen
  \bibfield  {author} {\bibinfo {author} {\bibfnamefont {G.}~\bibnamefont {Falasco}}, \bibinfo {author} {\bibfnamefont {R.}~\bibnamefont {Rao}},\ and\ \bibinfo {author} {\bibfnamefont {M.}~\bibnamefont {Esposito}},\ }\bibfield  {title} {\bibinfo {title} {Information thermodynamics of turing patterns},\ }\href@noop {} {\bibfield  {journal} {\bibinfo  {journal} {Physical Review Letters}\ }\textbf {\bibinfo {volume} {121}},\ \bibinfo {pages} {108301} (\bibinfo {year} {2018})}\BibitemShut {NoStop}%
\bibitem [{\citenamefont {Brauns}\ \emph {et~al.}(2020)\citenamefont {Brauns}, \citenamefont {Halatek},\ and\ \citenamefont {Frey}}]{brauns2020phase}%
  \BibitemOpen
  \bibfield  {author} {\bibinfo {author} {\bibfnamefont {F.}~\bibnamefont {Brauns}}, \bibinfo {author} {\bibfnamefont {J.}~\bibnamefont {Halatek}},\ and\ \bibinfo {author} {\bibfnamefont {E.}~\bibnamefont {Frey}},\ }\bibfield  {title} {\bibinfo {title} {Phase-space geometry of mass-conserving reaction-diffusion dynamics},\ }\href@noop {} {\bibfield  {journal} {\bibinfo  {journal} {Physical Review X}\ }\textbf {\bibinfo {volume} {10}},\ \bibinfo {pages} {041036} (\bibinfo {year} {2020})}\BibitemShut {NoStop}%
\bibitem [{\citenamefont {Busiello}\ \emph {et~al.}(2020)\citenamefont {Busiello}, \citenamefont {Gupta},\ and\ \citenamefont {Maritan}}]{busiello2020coarsegrained}%
  \BibitemOpen
  \bibfield  {author} {\bibinfo {author} {\bibfnamefont {D.~M.}\ \bibnamefont {Busiello}}, \bibinfo {author} {\bibfnamefont {D.}~\bibnamefont {Gupta}},\ and\ \bibinfo {author} {\bibfnamefont {A.}~\bibnamefont {Maritan}},\ }\bibfield  {title} {\bibinfo {title} {Coarse-grained entropy production with multiple reservoirs: Unraveling the role of time scales and detailed balance in biology-inspired systems},\ }\href@noop {} {\bibfield  {journal} {\bibinfo  {journal} {Physical Review Research}\ }\textbf {\bibinfo {volume} {2}},\ \bibinfo {pages} {043257} (\bibinfo {year} {2020})}\BibitemShut {NoStop}%
\bibitem [{\citenamefont {Nicoletti}\ and\ \citenamefont {Busiello}(2021)}]{nicoletti2021mutual}%
  \BibitemOpen
  \bibfield  {author} {\bibinfo {author} {\bibfnamefont {G.}~\bibnamefont {Nicoletti}}\ and\ \bibinfo {author} {\bibfnamefont {D.~M.}\ \bibnamefont {Busiello}},\ }\bibfield  {title} {\bibinfo {title} {Mutual information disentangles interactions from changing environments},\ }\href@noop {} {\bibfield  {journal} {\bibinfo  {journal} {Physical Review Letters}\ }\textbf {\bibinfo {volume} {127}},\ \bibinfo {pages} {228301} (\bibinfo {year} {2021})}\BibitemShut {NoStop}%
\bibitem [{\citenamefont {Nicoletti}\ and\ \citenamefont {Busiello}(2022)}]{nicoletti2022mutual}%
  \BibitemOpen
  \bibfield  {author} {\bibinfo {author} {\bibfnamefont {G.}~\bibnamefont {Nicoletti}}\ and\ \bibinfo {author} {\bibfnamefont {D.~M.}\ \bibnamefont {Busiello}},\ }\bibfield  {title} {\bibinfo {title} {Mutual information in changing environments: non-linear interactions, out-of-equilibrium systems, and continuously-varying diffusivities},\ }\href@noop {} {\bibfield  {journal} {\bibinfo  {journal} {Physical Review E}\ }\textbf {\bibinfo {volume} {106}},\ \bibinfo {pages} {014153} (\bibinfo {year} {2022})}\BibitemShut {NoStop}%
\bibitem [{\citenamefont {Elgeti}\ \emph {et~al.}(2015)\citenamefont {Elgeti}, \citenamefont {Winkler},\ and\ \citenamefont {Gompper}}]{elgeti2015physics}%
  \BibitemOpen
  \bibfield  {author} {\bibinfo {author} {\bibfnamefont {J.}~\bibnamefont {Elgeti}}, \bibinfo {author} {\bibfnamefont {R.~G.}\ \bibnamefont {Winkler}},\ and\ \bibinfo {author} {\bibfnamefont {G.}~\bibnamefont {Gompper}},\ }\bibfield  {title} {\bibinfo {title} {Physics of microswimmers—single particle motion and collective behavior: a review},\ }\href@noop {} {\bibfield  {journal} {\bibinfo  {journal} {Reports on progress in physics}\ }\textbf {\bibinfo {volume} {78}},\ \bibinfo {pages} {056601} (\bibinfo {year} {2015})}\BibitemShut {NoStop}%
\bibitem [{\citenamefont {Wu}\ and\ \citenamefont {Libchaber}(2000)}]{wu2000particle}%
  \BibitemOpen
  \bibfield  {author} {\bibinfo {author} {\bibfnamefont {X.-L.}\ \bibnamefont {Wu}}\ and\ \bibinfo {author} {\bibfnamefont {A.}~\bibnamefont {Libchaber}},\ }\bibfield  {title} {\bibinfo {title} {Particle diffusion in a quasi-two-dimensional bacterial bath},\ }\href@noop {} {\bibfield  {journal} {\bibinfo  {journal} {Physical review letters}\ }\textbf {\bibinfo {volume} {84}},\ \bibinfo {pages} {3017} (\bibinfo {year} {2000})}\BibitemShut {NoStop}%
\bibitem [{\citenamefont {Ghosh}\ \emph {et~al.}(2021)\citenamefont {Ghosh}, \citenamefont {Somasundar},\ and\ \citenamefont {Sen}}]{ghosh2021enzymes}%
  \BibitemOpen
  \bibfield  {author} {\bibinfo {author} {\bibfnamefont {S.}~\bibnamefont {Ghosh}}, \bibinfo {author} {\bibfnamefont {A.}~\bibnamefont {Somasundar}},\ and\ \bibinfo {author} {\bibfnamefont {A.}~\bibnamefont {Sen}},\ }\bibfield  {title} {\bibinfo {title} {Enzymes as active matter},\ }\href@noop {} {\bibfield  {journal} {\bibinfo  {journal} {Annual Review of Condensed Matter Physics}\ }\textbf {\bibinfo {volume} {12}},\ \bibinfo {pages} {177} (\bibinfo {year} {2021})}\BibitemShut {NoStop}%
\bibitem [{\citenamefont {Di~Terlizzi}\ \emph {et~al.}(2024)\citenamefont {Di~Terlizzi}, \citenamefont {Gironella}, \citenamefont {Herr{\'a}ez-Aguilar}, \citenamefont {Betz}, \citenamefont {Monroy}, \citenamefont {Baiesi},\ and\ \citenamefont {Ritort}}]{di2024variance}%
  \BibitemOpen
  \bibfield  {author} {\bibinfo {author} {\bibfnamefont {I.}~\bibnamefont {Di~Terlizzi}}, \bibinfo {author} {\bibfnamefont {M.}~\bibnamefont {Gironella}}, \bibinfo {author} {\bibfnamefont {D.}~\bibnamefont {Herr{\'a}ez-Aguilar}}, \bibinfo {author} {\bibfnamefont {T.}~\bibnamefont {Betz}}, \bibinfo {author} {\bibfnamefont {F.}~\bibnamefont {Monroy}}, \bibinfo {author} {\bibfnamefont {M.}~\bibnamefont {Baiesi}},\ and\ \bibinfo {author} {\bibfnamefont {F.}~\bibnamefont {Ritort}},\ }\bibfield  {title} {\bibinfo {title} {Variance sum rule for entropy production},\ }\href@noop {} {\bibfield  {journal} {\bibinfo  {journal} {Science}\ }\textbf {\bibinfo {volume} {383}},\ \bibinfo {pages} {971} (\bibinfo {year} {2024})}\BibitemShut {NoStop}%
\bibitem [{\citenamefont {Fodor}\ \emph {et~al.}(2022)\citenamefont {Fodor}, \citenamefont {Jack},\ and\ \citenamefont {Cates}}]{fodor2022irreversibility}%
  \BibitemOpen
  \bibfield  {author} {\bibinfo {author} {\bibfnamefont {{\'E}.}~\bibnamefont {Fodor}}, \bibinfo {author} {\bibfnamefont {R.~L.}\ \bibnamefont {Jack}},\ and\ \bibinfo {author} {\bibfnamefont {M.~E.}\ \bibnamefont {Cates}},\ }\bibfield  {title} {\bibinfo {title} {Irreversibility and biased ensembles in active matter: Insights from stochastic thermodynamics},\ }\href@noop {} {\bibfield  {journal} {\bibinfo  {journal} {Annual Review of Condensed Matter Physics}\ }\textbf {\bibinfo {volume} {13}},\ \bibinfo {pages} {215} (\bibinfo {year} {2022})}\BibitemShut {NoStop}%
\bibitem [{\citenamefont {J{\"u}licher}\ \emph {et~al.}(2018)\citenamefont {J{\"u}licher}, \citenamefont {Grill},\ and\ \citenamefont {Salbreux}}]{julicher2018hydrodynamic}%
  \BibitemOpen
  \bibfield  {author} {\bibinfo {author} {\bibfnamefont {F.}~\bibnamefont {J{\"u}licher}}, \bibinfo {author} {\bibfnamefont {S.~W.}\ \bibnamefont {Grill}},\ and\ \bibinfo {author} {\bibfnamefont {G.}~\bibnamefont {Salbreux}},\ }\bibfield  {title} {\bibinfo {title} {Hydrodynamic theory of active matter},\ }\href@noop {} {\bibfield  {journal} {\bibinfo  {journal} {Reports on Progress in Physics}\ }\textbf {\bibinfo {volume} {81}},\ \bibinfo {pages} {076601} (\bibinfo {year} {2018})}\BibitemShut {NoStop}%
\bibitem [{\citenamefont {Banerjee}\ \emph {et~al.}(2022)\citenamefont {Banerjee}, \citenamefont {Jack},\ and\ \citenamefont {Cates}}]{banerjee2022tracer}%
  \BibitemOpen
  \bibfield  {author} {\bibinfo {author} {\bibfnamefont {T.}~\bibnamefont {Banerjee}}, \bibinfo {author} {\bibfnamefont {R.~L.}\ \bibnamefont {Jack}},\ and\ \bibinfo {author} {\bibfnamefont {M.~E.}\ \bibnamefont {Cates}},\ }\bibfield  {title} {\bibinfo {title} {Tracer dynamics in one dimensional gases of active or passive particles},\ }\href@noop {} {\bibfield  {journal} {\bibinfo  {journal} {Journal of Statistical Mechanics: Theory and Experiment}\ }\textbf {\bibinfo {volume} {2022}},\ \bibinfo {pages} {013209} (\bibinfo {year} {2022})}\BibitemShut {NoStop}%
\bibitem [{\citenamefont {Xu}\ \emph {et~al.}(2019)\citenamefont {Xu}, \citenamefont {Ross}, \citenamefont {Valdez},\ and\ \citenamefont {Sen}}]{xu2019direct}%
  \BibitemOpen
  \bibfield  {author} {\bibinfo {author} {\bibfnamefont {M.}~\bibnamefont {Xu}}, \bibinfo {author} {\bibfnamefont {J.~L.}\ \bibnamefont {Ross}}, \bibinfo {author} {\bibfnamefont {L.}~\bibnamefont {Valdez}},\ and\ \bibinfo {author} {\bibfnamefont {A.}~\bibnamefont {Sen}},\ }\bibfield  {title} {\bibinfo {title} {Direct single molecule imaging of enhanced enzyme diffusion},\ }\href@noop {} {\bibfield  {journal} {\bibinfo  {journal} {Physical review letters}\ }\textbf {\bibinfo {volume} {123}},\ \bibinfo {pages} {128101} (\bibinfo {year} {2019})}\BibitemShut {NoStop}%
\bibitem [{\citenamefont {Maes}(2020)}]{maes2020fluctuating}%
  \BibitemOpen
  \bibfield  {author} {\bibinfo {author} {\bibfnamefont {C.}~\bibnamefont {Maes}},\ }\bibfield  {title} {\bibinfo {title} {Fluctuating motion in an active environment},\ }\href@noop {} {\bibfield  {journal} {\bibinfo  {journal} {Physical Review Letters}\ }\textbf {\bibinfo {volume} {125}},\ \bibinfo {pages} {208001} (\bibinfo {year} {2020})}\BibitemShut {NoStop}%
\bibitem [{\citenamefont {Dabelow}\ \emph {et~al.}(2019)\citenamefont {Dabelow}, \citenamefont {Bo},\ and\ \citenamefont {Eichhorn}}]{dabelow2019irreversibility}%
  \BibitemOpen
  \bibfield  {author} {\bibinfo {author} {\bibfnamefont {L.}~\bibnamefont {Dabelow}}, \bibinfo {author} {\bibfnamefont {S.}~\bibnamefont {Bo}},\ and\ \bibinfo {author} {\bibfnamefont {R.}~\bibnamefont {Eichhorn}},\ }\bibfield  {title} {\bibinfo {title} {Irreversibility in active matter systems: Fluctuation theorem and mutual information},\ }\href@noop {} {\bibfield  {journal} {\bibinfo  {journal} {Physical Review X}\ }\textbf {\bibinfo {volume} {9}},\ \bibinfo {pages} {021009} (\bibinfo {year} {2019})}\BibitemShut {NoStop}%
\bibitem [{\citenamefont {Hastings}(2010)}]{hastings2010timescales}%
  \BibitemOpen
  \bibfield  {author} {\bibinfo {author} {\bibfnamefont {A.}~\bibnamefont {Hastings}},\ }\bibfield  {title} {\bibinfo {title} {Timescales, dynamics, and ecological understanding},\ }\href@noop {} {\bibfield  {journal} {\bibinfo  {journal} {Ecology}\ }\textbf {\bibinfo {volume} {91}},\ \bibinfo {pages} {3471} (\bibinfo {year} {2010})}\BibitemShut {NoStop}%
\bibitem [{\citenamefont {Poisot}\ \emph {et~al.}(2015)\citenamefont {Poisot}, \citenamefont {Stouffer},\ and\ \citenamefont {Gravel}}]{poisot2015beyond}%
  \BibitemOpen
  \bibfield  {author} {\bibinfo {author} {\bibfnamefont {T.}~\bibnamefont {Poisot}}, \bibinfo {author} {\bibfnamefont {D.~B.}\ \bibnamefont {Stouffer}},\ and\ \bibinfo {author} {\bibfnamefont {D.}~\bibnamefont {Gravel}},\ }\bibfield  {title} {\bibinfo {title} {Beyond species: why ecological interaction networks vary through space and time},\ }\href@noop {} {\bibfield  {journal} {\bibinfo  {journal} {Oikos}\ }\textbf {\bibinfo {volume} {124}},\ \bibinfo {pages} {243} (\bibinfo {year} {2015})}\BibitemShut {NoStop}%
\bibitem [{\citenamefont {Timme}\ \emph {et~al.}(2014)\citenamefont {Timme}, \citenamefont {Ito}, \citenamefont {Myroshnychenko}, \citenamefont {Yeh}, \citenamefont {Hiolski}, \citenamefont {Hottowy},\ and\ \citenamefont {Beggs}}]{timme2014multiplex}%
  \BibitemOpen
  \bibfield  {author} {\bibinfo {author} {\bibfnamefont {N.}~\bibnamefont {Timme}}, \bibinfo {author} {\bibfnamefont {S.}~\bibnamefont {Ito}}, \bibinfo {author} {\bibfnamefont {M.}~\bibnamefont {Myroshnychenko}}, \bibinfo {author} {\bibfnamefont {F.-C.}\ \bibnamefont {Yeh}}, \bibinfo {author} {\bibfnamefont {E.}~\bibnamefont {Hiolski}}, \bibinfo {author} {\bibfnamefont {P.}~\bibnamefont {Hottowy}},\ and\ \bibinfo {author} {\bibfnamefont {J.~M.}\ \bibnamefont {Beggs}},\ }\bibfield  {title} {\bibinfo {title} {Multiplex networks of cortical and hippocampal neurons revealed at different timescales},\ }\href@noop {} {\bibfield  {journal} {\bibinfo  {journal} {PloS one}\ }\textbf {\bibinfo {volume} {9}},\ \bibinfo {pages} {e115764} (\bibinfo {year} {2014})}\BibitemShut {NoStop}%
\bibitem [{\citenamefont {Cavanagh}\ \emph {et~al.}(2020)\citenamefont {Cavanagh}, \citenamefont {Hunt},\ and\ \citenamefont {Kennerley}}]{cavanagh2020diversity}%
  \BibitemOpen
  \bibfield  {author} {\bibinfo {author} {\bibfnamefont {S.~E.}\ \bibnamefont {Cavanagh}}, \bibinfo {author} {\bibfnamefont {L.~T.}\ \bibnamefont {Hunt}},\ and\ \bibinfo {author} {\bibfnamefont {S.~W.}\ \bibnamefont {Kennerley}},\ }\bibfield  {title} {\bibinfo {title} {A diversity of intrinsic timescales underlie neural computations},\ }\href@noop {} {\bibfield  {journal} {\bibinfo  {journal} {Frontiers in Neural Circuits}\ }\textbf {\bibinfo {volume} {14}},\ \bibinfo {pages} {615626} (\bibinfo {year} {2020})}\BibitemShut {NoStop}%
\bibitem [{\citenamefont {Zeraati}\ \emph {et~al.}(2023)\citenamefont {Zeraati}, \citenamefont {Shi}, \citenamefont {Steinmetz}, \citenamefont {Gieselmann}, \citenamefont {Thiele}, \citenamefont {Moore}, \citenamefont {Levina},\ and\ \citenamefont {Engel}}]{zeraati2023intrinsic}%
  \BibitemOpen
  \bibfield  {author} {\bibinfo {author} {\bibfnamefont {R.}~\bibnamefont {Zeraati}}, \bibinfo {author} {\bibfnamefont {Y.-L.}\ \bibnamefont {Shi}}, \bibinfo {author} {\bibfnamefont {N.~A.}\ \bibnamefont {Steinmetz}}, \bibinfo {author} {\bibfnamefont {M.~A.}\ \bibnamefont {Gieselmann}}, \bibinfo {author} {\bibfnamefont {A.}~\bibnamefont {Thiele}}, \bibinfo {author} {\bibfnamefont {T.}~\bibnamefont {Moore}}, \bibinfo {author} {\bibfnamefont {A.}~\bibnamefont {Levina}},\ and\ \bibinfo {author} {\bibfnamefont {T.~A.}\ \bibnamefont {Engel}},\ }\bibfield  {title} {\bibinfo {title} {Intrinsic timescales in the visual cortex change with selective attention and reflect spatial connectivity},\ }\href@noop {} {\bibfield  {journal} {\bibinfo  {journal} {Nature communications}\ }\textbf {\bibinfo {volume} {14}},\ \bibinfo {pages} {1858} (\bibinfo {year} {2023})}\BibitemShut {NoStop}%
\bibitem [{\citenamefont {Nicoletti}\ \emph {et~al.}(2023{\natexlab{a}})\citenamefont {Nicoletti}, \citenamefont {Saravia}, \citenamefont {Momo}, \citenamefont {Maritan},\ and\ \citenamefont {Suweis}}]{nicoletti2023emergence}%
  \BibitemOpen
  \bibfield  {author} {\bibinfo {author} {\bibfnamefont {G.}~\bibnamefont {Nicoletti}}, \bibinfo {author} {\bibfnamefont {L.}~\bibnamefont {Saravia}}, \bibinfo {author} {\bibfnamefont {F.}~\bibnamefont {Momo}}, \bibinfo {author} {\bibfnamefont {A.}~\bibnamefont {Maritan}},\ and\ \bibinfo {author} {\bibfnamefont {S.}~\bibnamefont {Suweis}},\ }\bibfield  {title} {\bibinfo {title} {The emergence of scale-free fires in australia},\ }\href@noop {} {\bibfield  {journal} {\bibinfo  {journal} {Iscience}\ }\textbf {\bibinfo {volume} {26}} (\bibinfo {year} {2023}{\natexlab{a}})}\BibitemShut {NoStop}%
\bibitem [{\citenamefont {Nicoletti}\ and\ \citenamefont {Busiello}(2024{\natexlab{a}})}]{nicoletti2024information}%
  \BibitemOpen
  \bibfield  {author} {\bibinfo {author} {\bibfnamefont {G.}~\bibnamefont {Nicoletti}}\ and\ \bibinfo {author} {\bibfnamefont {D.~M.}\ \bibnamefont {Busiello}},\ }\bibfield  {title} {\bibinfo {title} {Information propagation in multilayer systems with higher-order interactions across timescales},\ }\href@noop {} {\bibfield  {journal} {\bibinfo  {journal} {Physical Review X}\ }\textbf {\bibinfo {volume} {14}},\ \bibinfo {pages} {021007} (\bibinfo {year} {2024}{\natexlab{a}})}\BibitemShut {NoStop}%
\bibitem [{\citenamefont {Nicoletti}\ and\ \citenamefont {Busiello}(2024{\natexlab{b}})}]{nicoletti2024gaussian}%
  \BibitemOpen
  \bibfield  {author} {\bibinfo {author} {\bibfnamefont {G.}~\bibnamefont {Nicoletti}}\ and\ \bibinfo {author} {\bibfnamefont {D.~M.}\ \bibnamefont {Busiello}},\ }\bibfield  {title} {\bibinfo {title} {Information propagation in gaussian processes on multilayer networks},\ }\href@noop {} {\bibfield  {journal} {\bibinfo  {journal} {arXiv preprint arXiv:2405.01363}\ } (\bibinfo {year} {2024}{\natexlab{b}})}\BibitemShut {NoStop}%
\bibitem [{\citenamefont {Barzon}\ \emph {et~al.}(2024)\citenamefont {Barzon}, \citenamefont {Busiello},\ and\ \citenamefont {Nicoletti}}]{barzon2024maximal}%
  \BibitemOpen
  \bibfield  {author} {\bibinfo {author} {\bibfnamefont {G.}~\bibnamefont {Barzon}}, \bibinfo {author} {\bibfnamefont {D.~M.}\ \bibnamefont {Busiello}},\ and\ \bibinfo {author} {\bibfnamefont {G.}~\bibnamefont {Nicoletti}},\ }\bibfield  {title} {\bibinfo {title} {Maximal information at the edge of stability in excitatory-inhibitory neural populations},\ }\href@noop {} {\bibfield  {journal} {\bibinfo  {journal} {arXiv preprint arXiv:2406.03380}\ } (\bibinfo {year} {2024})}\BibitemShut {NoStop}%
\bibitem [{\citenamefont {Sharma}\ \emph {et~al.}(2017)\citenamefont {Sharma}, \citenamefont {Sharma},\ and\ \citenamefont {Athaiya}}]{sharma2017activation}%
  \BibitemOpen
  \bibfield  {author} {\bibinfo {author} {\bibfnamefont {S.}~\bibnamefont {Sharma}}, \bibinfo {author} {\bibfnamefont {S.}~\bibnamefont {Sharma}},\ and\ \bibinfo {author} {\bibfnamefont {A.}~\bibnamefont {Athaiya}},\ }\bibfield  {title} {\bibinfo {title} {Activation functions in neural networks},\ }\href@noop {} {\bibfield  {journal} {\bibinfo  {journal} {Towards Data Sci}\ }\textbf {\bibinfo {volume} {6}},\ \bibinfo {pages} {310} (\bibinfo {year} {2017})}\BibitemShut {NoStop}%
\bibitem [{\citenamefont {Christodoulou}\ \emph {et~al.}(2022)\citenamefont {Christodoulou}, \citenamefont {Vogels},\ and\ \citenamefont {Agnes}}]{christodoulou2022regimes}%
  \BibitemOpen
  \bibfield  {author} {\bibinfo {author} {\bibfnamefont {G.}~\bibnamefont {Christodoulou}}, \bibinfo {author} {\bibfnamefont {T.~P.}\ \bibnamefont {Vogels}},\ and\ \bibinfo {author} {\bibfnamefont {E.~J.}\ \bibnamefont {Agnes}},\ }\bibfield  {title} {\bibinfo {title} {Regimes and mechanisms of transient amplification in abstract and biological neural networks},\ }\href@noop {} {\bibfield  {journal} {\bibinfo  {journal} {PLoS Computational Biology}\ }\textbf {\bibinfo {volume} {18}},\ \bibinfo {pages} {e1010365} (\bibinfo {year} {2022})}\BibitemShut {NoStop}%
\bibitem [{\citenamefont {Kunc}\ and\ \citenamefont {Kl{\'e}ma}(2021)}]{kunc2021transformative}%
  \BibitemOpen
  \bibfield  {author} {\bibinfo {author} {\bibfnamefont {V.}~\bibnamefont {Kunc}}\ and\ \bibinfo {author} {\bibfnamefont {J.}~\bibnamefont {Kl{\'e}ma}},\ }\bibfield  {title} {\bibinfo {title} {On transformative adaptive activation functions in neural networks for gene expression inference},\ }\href@noop {} {\bibfield  {journal} {\bibinfo  {journal} {Plos one}\ }\textbf {\bibinfo {volume} {16}},\ \bibinfo {pages} {e0243915} (\bibinfo {year} {2021})}\BibitemShut {NoStop}%
\bibitem [{\citenamefont {Kraskov}\ \emph {et~al.}(2004)\citenamefont {Kraskov}, \citenamefont {St{\"o}gbauer},\ and\ \citenamefont {Grassberger}}]{kraskov2004estimating}%
  \BibitemOpen
  \bibfield  {author} {\bibinfo {author} {\bibfnamefont {A.}~\bibnamefont {Kraskov}}, \bibinfo {author} {\bibfnamefont {H.}~\bibnamefont {St{\"o}gbauer}},\ and\ \bibinfo {author} {\bibfnamefont {P.}~\bibnamefont {Grassberger}},\ }\bibfield  {title} {\bibinfo {title} {Estimating mutual information},\ }\href@noop {} {\bibfield  {journal} {\bibinfo  {journal} {Physical Review E—Statistical, Nonlinear, and Soft Matter Physics}\ }\textbf {\bibinfo {volume} {69}},\ \bibinfo {pages} {066138} (\bibinfo {year} {2004})}\BibitemShut {NoStop}%
\bibitem [{\citenamefont {Nicoletti}\ and\ \citenamefont {Busiello}(2024{\natexlab{c}})}]{nicoletti2024tuning}%
  \BibitemOpen
  \bibfield  {author} {\bibinfo {author} {\bibfnamefont {G.}~\bibnamefont {Nicoletti}}\ and\ \bibinfo {author} {\bibfnamefont {D.~M.}\ \bibnamefont {Busiello}},\ }\bibfield  {title} {\bibinfo {title} {Tuning transduction from hidden observables to optimize information harvesting},\ }\href@noop {} {\bibfield  {journal} {\bibinfo  {journal} {Physical Review Letters}\ } (\bibinfo {year} {in press, 2024}{\natexlab{c}})}\BibitemShut {NoStop}%
\bibitem [{\citenamefont {Wang}\ \emph {et~al.}(2020)\citenamefont {Wang}, \citenamefont {Seno}, \citenamefont {Sokolov}, \citenamefont {Chechkin},\ and\ \citenamefont {Metzler}}]{wang2020unexpected}%
  \BibitemOpen
  \bibfield  {author} {\bibinfo {author} {\bibfnamefont {W.}~\bibnamefont {Wang}}, \bibinfo {author} {\bibfnamefont {F.}~\bibnamefont {Seno}}, \bibinfo {author} {\bibfnamefont {I.~M.}\ \bibnamefont {Sokolov}}, \bibinfo {author} {\bibfnamefont {A.~V.}\ \bibnamefont {Chechkin}},\ and\ \bibinfo {author} {\bibfnamefont {R.}~\bibnamefont {Metzler}},\ }\bibfield  {title} {\bibinfo {title} {Unexpected crossovers in correlated random-diffusivity processes},\ }\href {https://doi.org/10.1088/1367-2630/aba390} {\bibfield  {journal} {\bibinfo  {journal} {New Journal of Physics}\ }\textbf {\bibinfo {volume} {22}},\ \bibinfo {pages} {083041} (\bibinfo {year} {2020})}\BibitemShut {NoStop}%
\bibitem [{\citenamefont {Mariani}\ \emph {et~al.}(2022)\citenamefont {Mariani}, \citenamefont {Nicoletti}, \citenamefont {Bisio}, \citenamefont {Maschietto}, \citenamefont {Vassanelli},\ and\ \citenamefont {Suweis}}]{mariani2022disentangling}%
  \BibitemOpen
  \bibfield  {author} {\bibinfo {author} {\bibfnamefont {B.}~\bibnamefont {Mariani}}, \bibinfo {author} {\bibfnamefont {G.}~\bibnamefont {Nicoletti}}, \bibinfo {author} {\bibfnamefont {M.}~\bibnamefont {Bisio}}, \bibinfo {author} {\bibfnamefont {M.}~\bibnamefont {Maschietto}}, \bibinfo {author} {\bibfnamefont {S.}~\bibnamefont {Vassanelli}},\ and\ \bibinfo {author} {\bibfnamefont {S.}~\bibnamefont {Suweis}},\ }\bibfield  {title} {\bibinfo {title} {Disentangling the critical signatures of neural activity},\ }\href@noop {} {\bibfield  {journal} {\bibinfo  {journal} {Scientific reports}\ }\textbf {\bibinfo {volume} {12}},\ \bibinfo {pages} {10770} (\bibinfo {year} {2022})}\BibitemShut {NoStop}%
\bibitem [{\citenamefont {Honey}\ \emph {et~al.}(2017)\citenamefont {Honey}, \citenamefont {Newman},\ and\ \citenamefont {Schapiro}}]{honey2017switching}%
  \BibitemOpen
  \bibfield  {author} {\bibinfo {author} {\bibfnamefont {C.~J.}\ \bibnamefont {Honey}}, \bibinfo {author} {\bibfnamefont {E.~L.}\ \bibnamefont {Newman}},\ and\ \bibinfo {author} {\bibfnamefont {A.~C.}\ \bibnamefont {Schapiro}},\ }\bibfield  {title} {\bibinfo {title} {Switching between internal and external modes: A multiscale learning principle},\ }\href@noop {} {\bibfield  {journal} {\bibinfo  {journal} {Network Neuroscience}\ }\textbf {\bibinfo {volume} {1}},\ \bibinfo {pages} {339} (\bibinfo {year} {2017})}\BibitemShut {NoStop}%
\bibitem [{\citenamefont {Nicoletti}\ \emph {et~al.}(2020)\citenamefont {Nicoletti}, \citenamefont {Suweis},\ and\ \citenamefont {Maritan}}]{nicoletti2020scaling}%
  \BibitemOpen
  \bibfield  {author} {\bibinfo {author} {\bibfnamefont {G.}~\bibnamefont {Nicoletti}}, \bibinfo {author} {\bibfnamefont {S.}~\bibnamefont {Suweis}},\ and\ \bibinfo {author} {\bibfnamefont {A.}~\bibnamefont {Maritan}},\ }\bibfield  {title} {\bibinfo {title} {Scaling and criticality in a phenomenological renormalization group},\ }\href@noop {} {\bibfield  {journal} {\bibinfo  {journal} {Physical Review Research}\ }\textbf {\bibinfo {volume} {2}},\ \bibinfo {pages} {023144} (\bibinfo {year} {2020})}\BibitemShut {NoStop}%
\bibitem [{\citenamefont {Nicoletti}\ \emph {et~al.}(2023{\natexlab{b}})\citenamefont {Nicoletti}, \citenamefont {Padmanabha}, \citenamefont {Azaele}, \citenamefont {Suweis}, \citenamefont {Rinaldo},\ and\ \citenamefont {Maritan}}]{nicoletti2023emergent}%
  \BibitemOpen
  \bibfield  {author} {\bibinfo {author} {\bibfnamefont {G.}~\bibnamefont {Nicoletti}}, \bibinfo {author} {\bibfnamefont {P.}~\bibnamefont {Padmanabha}}, \bibinfo {author} {\bibfnamefont {S.}~\bibnamefont {Azaele}}, \bibinfo {author} {\bibfnamefont {S.}~\bibnamefont {Suweis}}, \bibinfo {author} {\bibfnamefont {A.}~\bibnamefont {Rinaldo}},\ and\ \bibinfo {author} {\bibfnamefont {A.}~\bibnamefont {Maritan}},\ }\bibfield  {title} {\bibinfo {title} {Emergent encoding of dispersal network topologies in spatial metapopulation models},\ }\href@noop {} {\bibfield  {journal} {\bibinfo  {journal} {Proceedings of the National Academy of Sciences}\ }\textbf {\bibinfo {volume} {120}},\ \bibinfo {pages} {e2311548120} (\bibinfo {year} {2023}{\natexlab{b}})}\BibitemShut {NoStop}%
\bibitem [{\citenamefont {Ferrari}\ \emph {et~al.}(2018)\citenamefont {Ferrari}, \citenamefont {Deny}, \citenamefont {Chalk}, \citenamefont {Tka{\v{c}}ik}, \citenamefont {Marre},\ and\ \citenamefont {Mora}}]{ferrari2018separating}%
  \BibitemOpen
  \bibfield  {author} {\bibinfo {author} {\bibfnamefont {U.}~\bibnamefont {Ferrari}}, \bibinfo {author} {\bibfnamefont {S.}~\bibnamefont {Deny}}, \bibinfo {author} {\bibfnamefont {M.}~\bibnamefont {Chalk}}, \bibinfo {author} {\bibfnamefont {G.}~\bibnamefont {Tka{\v{c}}ik}}, \bibinfo {author} {\bibfnamefont {O.}~\bibnamefont {Marre}},\ and\ \bibinfo {author} {\bibfnamefont {T.}~\bibnamefont {Mora}},\ }\bibfield  {title} {\bibinfo {title} {Separating intrinsic interactions from extrinsic correlations in a network of sensory neurons},\ }\href@noop {} {\bibfield  {journal} {\bibinfo  {journal} {Physical Review E}\ }\textbf {\bibinfo {volume} {98}},\ \bibinfo {pages} {042410} (\bibinfo {year} {2018})}\BibitemShut {NoStop}%
\bibitem [{\citenamefont {Sporns}(2011)}]{sporns2011human}%
  \BibitemOpen
  \bibfield  {author} {\bibinfo {author} {\bibfnamefont {O.}~\bibnamefont {Sporns}},\ }\bibfield  {title} {\bibinfo {title} {The human connectome: a complex network},\ }\href@noop {} {\bibfield  {journal} {\bibinfo  {journal} {Annals of the new York Academy of Sciences}\ }\textbf {\bibinfo {volume} {1224}},\ \bibinfo {pages} {109} (\bibinfo {year} {2011})}\BibitemShut {NoStop}%
\bibitem [{\citenamefont {Mi{\v{s}}i{\'c}}\ \emph {et~al.}(2015)\citenamefont {Mi{\v{s}}i{\'c}}, \citenamefont {Betzel}, \citenamefont {Nematzadeh}, \citenamefont {Goni}, \citenamefont {Griffa}, \citenamefont {Hagmann}, \citenamefont {Flammini}, \citenamefont {Ahn},\ and\ \citenamefont {Sporns}}]{mivsic2015cooperative}%
  \BibitemOpen
  \bibfield  {author} {\bibinfo {author} {\bibfnamefont {B.}~\bibnamefont {Mi{\v{s}}i{\'c}}}, \bibinfo {author} {\bibfnamefont {R.~F.}\ \bibnamefont {Betzel}}, \bibinfo {author} {\bibfnamefont {A.}~\bibnamefont {Nematzadeh}}, \bibinfo {author} {\bibfnamefont {J.}~\bibnamefont {Goni}}, \bibinfo {author} {\bibfnamefont {A.}~\bibnamefont {Griffa}}, \bibinfo {author} {\bibfnamefont {P.}~\bibnamefont {Hagmann}}, \bibinfo {author} {\bibfnamefont {A.}~\bibnamefont {Flammini}}, \bibinfo {author} {\bibfnamefont {Y.-Y.}\ \bibnamefont {Ahn}},\ and\ \bibinfo {author} {\bibfnamefont {O.}~\bibnamefont {Sporns}},\ }\bibfield  {title} {\bibinfo {title} {Cooperative and competitive spreading dynamics on the human connectome},\ }\href@noop {} {\bibfield  {journal} {\bibinfo  {journal} {Neuron}\ }\textbf {\bibinfo {volume} {86}},\ \bibinfo {pages} {1518} (\bibinfo {year} {2015})}\BibitemShut {NoStop}%
\bibitem [{\citenamefont {Barzon}\ \emph {et~al.}(2022)\citenamefont {Barzon}, \citenamefont {Nicoletti}, \citenamefont {Mariani}, \citenamefont {Formentin},\ and\ \citenamefont {Suweis}}]{barzon2022criticality}%
  \BibitemOpen
  \bibfield  {author} {\bibinfo {author} {\bibfnamefont {G.}~\bibnamefont {Barzon}}, \bibinfo {author} {\bibfnamefont {G.}~\bibnamefont {Nicoletti}}, \bibinfo {author} {\bibfnamefont {B.}~\bibnamefont {Mariani}}, \bibinfo {author} {\bibfnamefont {M.}~\bibnamefont {Formentin}},\ and\ \bibinfo {author} {\bibfnamefont {S.}~\bibnamefont {Suweis}},\ }\bibfield  {title} {\bibinfo {title} {Criticality and network structure drive emergent oscillations in a stochastic whole-brain model},\ }\href@noop {} {\bibfield  {journal} {\bibinfo  {journal} {Journal of Physics: Complexity}\ }\textbf {\bibinfo {volume} {3}},\ \bibinfo {pages} {025010} (\bibinfo {year} {2022})}\BibitemShut {NoStop}%
\bibitem [{\citenamefont {Stepanyants}\ \emph {et~al.}(2002)\citenamefont {Stepanyants}, \citenamefont {Hof},\ and\ \citenamefont {Chklovskii}}]{stepanyants2002geometry}%
  \BibitemOpen
  \bibfield  {author} {\bibinfo {author} {\bibfnamefont {A.}~\bibnamefont {Stepanyants}}, \bibinfo {author} {\bibfnamefont {P.~R.}\ \bibnamefont {Hof}},\ and\ \bibinfo {author} {\bibfnamefont {D.~B.}\ \bibnamefont {Chklovskii}},\ }\bibfield  {title} {\bibinfo {title} {Geometry and structural plasticity of synaptic connectivity},\ }\href@noop {} {\bibfield  {journal} {\bibinfo  {journal} {Neuron}\ }\textbf {\bibinfo {volume} {34}},\ \bibinfo {pages} {275} (\bibinfo {year} {2002})}\BibitemShut {NoStop}%
\bibitem [{\citenamefont {Stampanoni~Bassi}\ \emph {et~al.}(2019)\citenamefont {Stampanoni~Bassi}, \citenamefont {Iezzi}, \citenamefont {Gilio}, \citenamefont {Centonze},\ and\ \citenamefont {Buttari}}]{stampanoni2019synaptic}%
  \BibitemOpen
  \bibfield  {author} {\bibinfo {author} {\bibfnamefont {M.}~\bibnamefont {Stampanoni~Bassi}}, \bibinfo {author} {\bibfnamefont {E.}~\bibnamefont {Iezzi}}, \bibinfo {author} {\bibfnamefont {L.}~\bibnamefont {Gilio}}, \bibinfo {author} {\bibfnamefont {D.}~\bibnamefont {Centonze}},\ and\ \bibinfo {author} {\bibfnamefont {F.}~\bibnamefont {Buttari}},\ }\bibfield  {title} {\bibinfo {title} {Synaptic plasticity shapes brain connectivity: implications for network topology},\ }\href@noop {} {\bibfield  {journal} {\bibinfo  {journal} {International journal of molecular sciences}\ }\textbf {\bibinfo {volume} {20}},\ \bibinfo {pages} {6193} (\bibinfo {year} {2019})}\BibitemShut {NoStop}%
\bibitem [{\citenamefont {Schaffer}\ and\ \citenamefont {Ideker}(2021)}]{schaffer2021mapping}%
  \BibitemOpen
  \bibfield  {author} {\bibinfo {author} {\bibfnamefont {L.~V.}\ \bibnamefont {Schaffer}}\ and\ \bibinfo {author} {\bibfnamefont {T.}~\bibnamefont {Ideker}},\ }\bibfield  {title} {\bibinfo {title} {Mapping the multiscale structure of biological systems},\ }\href@noop {} {\bibfield  {journal} {\bibinfo  {journal} {Cell systems}\ }\textbf {\bibinfo {volume} {12}},\ \bibinfo {pages} {622} (\bibinfo {year} {2021})}\BibitemShut {NoStop}%
\bibitem [{\citenamefont {Boccaletti}\ \emph {et~al.}(2014)\citenamefont {Boccaletti}, \citenamefont {Bianconi}, \citenamefont {Criado}, \citenamefont {Del~Genio}, \citenamefont {G{\'o}mez-Gardenes}, \citenamefont {Romance}, \citenamefont {Sendina-Nadal}, \citenamefont {Wang},\ and\ \citenamefont {Zanin}}]{boccaletti2014structure}%
  \BibitemOpen
  \bibfield  {author} {\bibinfo {author} {\bibfnamefont {S.}~\bibnamefont {Boccaletti}}, \bibinfo {author} {\bibfnamefont {G.}~\bibnamefont {Bianconi}}, \bibinfo {author} {\bibfnamefont {R.}~\bibnamefont {Criado}}, \bibinfo {author} {\bibfnamefont {C.~I.}\ \bibnamefont {Del~Genio}}, \bibinfo {author} {\bibfnamefont {J.}~\bibnamefont {G{\'o}mez-Gardenes}}, \bibinfo {author} {\bibfnamefont {M.}~\bibnamefont {Romance}}, \bibinfo {author} {\bibfnamefont {I.}~\bibnamefont {Sendina-Nadal}}, \bibinfo {author} {\bibfnamefont {Z.}~\bibnamefont {Wang}},\ and\ \bibinfo {author} {\bibfnamefont {M.}~\bibnamefont {Zanin}},\ }\bibfield  {title} {\bibinfo {title} {The structure and dynamics of multilayer networks},\ }\href@noop {} {\bibfield  {journal} {\bibinfo  {journal} {Physics reports}\ }\textbf {\bibinfo {volume} {544}},\ \bibinfo {pages} {1} (\bibinfo {year} {2014})}\BibitemShut {NoStop}%
\bibitem [{\citenamefont {Bianconi}(2018)}]{bianconi2018multilayer}%
  \BibitemOpen
  \bibfield  {author} {\bibinfo {author} {\bibfnamefont {G.}~\bibnamefont {Bianconi}},\ }\href@noop {} {\emph {\bibinfo {title} {Multilayer networks: structure and function}}}\ (\bibinfo  {publisher} {Oxford university press},\ \bibinfo {year} {2018})\BibitemShut {NoStop}%
\bibitem [{\citenamefont {Aleta}\ and\ \citenamefont {Moreno}(2019)}]{aleta2019multilayer}%
  \BibitemOpen
  \bibfield  {author} {\bibinfo {author} {\bibfnamefont {A.}~\bibnamefont {Aleta}}\ and\ \bibinfo {author} {\bibfnamefont {Y.}~\bibnamefont {Moreno}},\ }\bibfield  {title} {\bibinfo {title} {Multilayer networks in a nutshell},\ }\href@noop {} {\bibfield  {journal} {\bibinfo  {journal} {Annual Review of Condensed Matter Physics}\ }\textbf {\bibinfo {volume} {10}},\ \bibinfo {pages} {45} (\bibinfo {year} {2019})}\BibitemShut {NoStop}%
\bibitem [{\citenamefont {Artime}\ \emph {et~al.}(2022)\citenamefont {Artime}, \citenamefont {Benigni}, \citenamefont {Bertagnolli}, \citenamefont {d'Andrea}, \citenamefont {Gallotti}, \citenamefont {Ghavasieh}, \citenamefont {Raimondo},\ and\ \citenamefont {De~Domenico}}]{artime2022multilayer}%
  \BibitemOpen
  \bibfield  {author} {\bibinfo {author} {\bibfnamefont {O.}~\bibnamefont {Artime}}, \bibinfo {author} {\bibfnamefont {B.}~\bibnamefont {Benigni}}, \bibinfo {author} {\bibfnamefont {G.}~\bibnamefont {Bertagnolli}}, \bibinfo {author} {\bibfnamefont {V.}~\bibnamefont {d'Andrea}}, \bibinfo {author} {\bibfnamefont {R.}~\bibnamefont {Gallotti}}, \bibinfo {author} {\bibfnamefont {A.}~\bibnamefont {Ghavasieh}}, \bibinfo {author} {\bibfnamefont {S.}~\bibnamefont {Raimondo}},\ and\ \bibinfo {author} {\bibfnamefont {M.}~\bibnamefont {De~Domenico}},\ }\href@noop {} {\emph {\bibinfo {title} {Multilayer network science: from cells to societies}}}\ (\bibinfo  {publisher} {Cambridge University Press},\ \bibinfo {year} {2022})\BibitemShut {NoStop}%
\bibitem [{\citenamefont {Parrondo}\ \emph {et~al.}(2015)\citenamefont {Parrondo}, \citenamefont {Horowitz},\ and\ \citenamefont {Sagawa}}]{parrondo2015thermodynamics}%
  \BibitemOpen
  \bibfield  {author} {\bibinfo {author} {\bibfnamefont {J.~M.}\ \bibnamefont {Parrondo}}, \bibinfo {author} {\bibfnamefont {J.~M.}\ \bibnamefont {Horowitz}},\ and\ \bibinfo {author} {\bibfnamefont {T.}~\bibnamefont {Sagawa}},\ }\bibfield  {title} {\bibinfo {title} {Thermodynamics of information},\ }\href@noop {} {\bibfield  {journal} {\bibinfo  {journal} {Nature physics}\ }\textbf {\bibinfo {volume} {11}},\ \bibinfo {pages} {131} (\bibinfo {year} {2015})}\BibitemShut {NoStop}%
\bibitem [{\citenamefont {Sagawa}(2012)}]{sagawa2012thermodynamics}%
  \BibitemOpen
  \bibfield  {author} {\bibinfo {author} {\bibfnamefont {T.}~\bibnamefont {Sagawa}},\ }\bibfield  {title} {\bibinfo {title} {Thermodynamics of information processing in small systems},\ }\href@noop {} {\bibfield  {journal} {\bibinfo  {journal} {Progress of theoretical physics}\ }\textbf {\bibinfo {volume} {127}},\ \bibinfo {pages} {1} (\bibinfo {year} {2012})}\BibitemShut {NoStop}%
\bibitem [{\citenamefont {Horowitz}\ and\ \citenamefont {Esposito}(2014)}]{horowitz2014thermodynamics}%
  \BibitemOpen
  \bibfield  {author} {\bibinfo {author} {\bibfnamefont {J.~M.}\ \bibnamefont {Horowitz}}\ and\ \bibinfo {author} {\bibfnamefont {M.}~\bibnamefont {Esposito}},\ }\bibfield  {title} {\bibinfo {title} {Thermodynamics with continuous information flow},\ }\href@noop {} {\bibfield  {journal} {\bibinfo  {journal} {Physical Review X}\ }\textbf {\bibinfo {volume} {4}},\ \bibinfo {pages} {031015} (\bibinfo {year} {2014})}\BibitemShut {NoStop}%
\end{thebibliography}

%

\end{document}